\documentclass[10pt]{article}

\usepackage{geometry}
\usepackage{float}
\usepackage{url,hyperref}

\usepackage{graphicx}
\usepackage{multirow}
\usepackage{amsmath,amssymb,amsfonts}
\usepackage{amsthm}
\usepackage{mathrsfs}
\usepackage[title]{appendix}
\usepackage{xcolor}
\usepackage{textcomp}
\usepackage{manyfoot}
\usepackage{booktabs}
\usepackage{algorithm}
\usepackage{algorithmicx}
\usepackage{algpseudocode}
\usepackage{listings}
\usepackage[utf8]{inputenc}
\usepackage{bigints}                       
\usepackage{mathtools}  
                      
\usepackage{subcaption}                    
\usepackage{enumitem}                      
\usepackage{physics}                       
\usepackage{esdiff} 
 
\usepackage{quoting}   
\usepackage{gensymb} 

\usepackage{dcolumn}
\usepackage{bm}
\usepackage[T1]{fontenc}
\usepackage{mathptmx}
\usepackage[bottom]{footmisc}
\usepackage{wasysym}
\usepackage[angle]{natbib}
\usepackage{doi}
\usepackage{comment}

\renewcommand{\epsilon}{\varepsilon} 
\renewcommand{\theta}{\vartheta}     
\renewcommand{\kappa}{\varkappa}     
\renewcommand{\rho}{\varrho}         
\renewcommand{\phi}{\varphi}         

\newcommand\leftidx[3]{%
	{\vphantom{#2}}#1#2#3%
}
\DeclareMathOperator\sign{sgn}       

\newcommand*{\setdef}[1]{
	\begingroup
	\def\or{\textup{ or }}%
	\def\and{,\ }%
	\left\{#1 \right\}%
	\endgroup
} 

\begin{document}

\title{Three ways to find comfort with the Bell proof and the results of the Bell experiments}

\author{Richard D. Gill, University of Leiden\\Inge S. Helland, University of Oslo \\ Bart Jongejan, University of Copenhagen} 

\date{}

\maketitle

\begin{abstract}
Bell's theorem states that no description of a Bell experiment can be simultaneously local, realistic in the sense of counterfactual definiteness, and free of conspiracy between settings and hidden state.
The recent generation of experiments has confirmed the predicted violation of the CHSH inequality, so one of the assumptions must be abandoned.
Which one, and how one reconstructs a coherent worldview after doing so, is a question on which many authors disagree.
This paper is written by three such authors.
All three reject counterfactual definiteness. Two of us reject violation of the statistical independence condition of setting choices and state, by many authors termed the ``no conspiracy condition''. The third accepts a violation of this condition in his model, but argues that it involves no conspiracy.
After a joint exposition of the classical half of Bell's theorem in the language of Pearl-style causal graphs, a joint summary of the loophole-free experiments, and a joint survey of the recent literature, each author states where they have presently arrived.
Richard Gill accepts irreducible and non-local quantum randomness and finds the choice between locality and realism a false dichotomy.
In his later works, Bell derives counterfactual definiteness from classical local causality, and that is what has to go.
The metaphysical concepts ``realism'', ``locality'', ``causality'' need to be reconsidered.
Inge Helland reconstructs the Hilbert-space formalism from a theory of accessible variables, and from this theory he concludes that every observer must be limited in a specific sense.
Bart Jongejan proposes a geometric hidden-variable construction in which the degree of violation of the CHSH inequality depends on the number of dimensions of space, Tsirelson's bound corresponding to three dimensions.
In this construction, each party's hidden variable is distributed independently of the distant party's setting, but the joint distribution of the two hidden variables depends on both settings, so the statistical independence condition is not satisfied. His model is no-signalling but, like QM, not locally causal.
The authors conclude with a discussion.
\end{abstract}


\section{Introduction}\label{Ch1}

More than sixty years after \citet{PhysicsPhysiqueFizika.1.195},
and more than ten years after the first loophole-free experiments,
the empirical situation is very clear: the Bell-CHSH inequality is violated in the laboratory,
by the amount predicted by quantum mechanics,
under conditions that close the loopholes of locality and detection.
There is some disagreement about whether also the freedom-of-choice loophole is closed. Some groups behind the experiments have said it is, whereas others don't claim that or even say that it is in principle not possible to close the freedom-of-choice loophole.
The Bell theorem says that any description of such an experiment that is at once local,
realistic (in the sense of counterfactual definiteness),
and free of conspiracy between settings and hidden states is mathematically inconsistent with those correlations.
At least one of the three must go.
The experiments do not tell us which.

The experiments also do not tell us how to live with the choice.
Different readers of the same proof and the same data arrive at different conclusions,
and the difference is not a matter of who has done the mathematics correctly -- we agree on the mathematics -- but of which background picture of the world one is willing to give up,
and in exchange for what.
The present paper started from an email correspondence among three authors who agree on Section~\ref{Ch2} of what follows,
agree on the experimental record of Section \ref{Ch3},
and nevertheless arrive at different interpretations.
Rather than hide the disagreement,
we have chosen to make it the subject of the paper.

The structure of the paper is as follows.
Section~\ref{Ch2} states the classical half of Bell's theorem in the language of Pearl-style causal graphs: it is here that the assumptions are made precise,
and the CHSH inequality is derived.
Section \ref{Ch3} summarises the Bell experiments and the different loopholes that the modern generation of experiments has closed.
A reader who already accepts that the CHSH inequality has been empirically violated may skim this section.
Section \ref{Ch4} discusses the relevant recent literature,
including the superdeterministic,
non-local,
and Bell-denying responses that we do not adopt,
and gives reasons why we find none of them a comfortable resting place.
Section \ref{RiHeBa} is the heart of the paper.
Each of the three authors states,
in the first person,
where the argument of Sections \ref{Ch2},
\ref{Ch3} and \ref{Ch4} leaves him.

Richard Gill (Sec. \ref{Richard}) accepts that the randomness one sees in a Bell experiment is irreducible and non-local.
He follows an approach pioneered by V.P.~Belavkin in which a Heisenberg cut is placed between past and future instead of between microscopic and macroscopic.
Belavkin called his approach ``eventum mechanics'' and saw it as a mathematical solution to the so-called measurement problem or Schr\"odinger cat problem.
The cut is relative to a point in space-time,
it is relative to ``time and place''.
The ontology is that what has happened in the past light-cone of a point in space-time is real.
The quantum state here and now is the quantum state of the future as seen from here and now.
As the arrow of time progresses the quantum wave of the future is continually collapsing at the boundary between past and future,
past reality is enlarged at this boundary,
and the quantum state of the future conditional on the past is updated.
The past as seen by different travellers in space-time (always moving below the speed of light) is the same where the paths intersect
and entitled to be considered as real.
It doesn't matter whether the travellers are conscious beings or automatons.
Quantum mechanics hereby defines the evolution of a real past as a stochastic process through time.
Accepting this world picture entails re-evaluating the concepts of realism,
locality and causality.
It does not require the mind of an observer.
Quantum probability is physical,
it is not a consequence of lack of information.
This unification of the deterministic and stochastic parts of quantum mechanics is a stochastic model,
not a deterministic model.

Inge Helland (Sec. \ref{Inge}) gives up realism in the sense that most people define this term.
His starting point is a rebuilding of the Hilbert-space formalism of quantum mechanics from a theory of theoretical variables that may be accessible or non-accessible.
His first theorem assumes a situation with two complementary (non-equivalent maximal accessible) variables,
and from this he proves that all accessible variables in this situation have suitable self-adjoint operators associated with them.
Another mathematical theorem gives conditions under which a given actor is unable to keep enough variables during a modelling process. This is built upon the concepts of related and non-related maximal accessible variables.
From this,
he derives a general argument against counterfactual definiteness, connected to the Bell experiment, but not directly to the Bell theorem.
Inge's interpretation of quantum mechanics is epistemic,
connected to an observer/actor or a communicating group of actors.
It is argued that in cases like this,
where the actual actor is arbitrary,
other communicating people will tend to agree with the arguments that the
CHSH inequality cannot be applied, implying that not all assumptions underlying its derivation can be valid.

Bart Jongejan (Sec.~\ref{Bart}) regards the rejection of counterfactual definiteness as the basis for a hidden-variable model that recovers the violation of the CHSH inequality. He argues that the degree of violation is linked to the dimensionality of space. Another peculiar aspect of Bart's model is that the geometric relations between the setting directions of two widely separated spin component measuring apparatuses are seen as the result of a Bell-type experiment with many trials, not as \textit{a priori} given inputs. 
The model shows that the coordination between two widely separated measurements can be described by hidden variables $\lambda_A$ and $\lambda_B$ that, over many trials, are correlated with settings $A$ and $B$ in just the right way to violate the CHSH inequality, and also such that $\lambda_A$ is independent of $B$ and $\lambda_B$ is independent of $A$. The hidden-variable model rejects the probability space underlying Bell's theorem; there is no single hidden variable whose distribution is the same for all setting pairs. The model violates statistical independence and, since it involves neither superdeterminism nor retrocausality, also local causality, like QM: the joint distribution of the two hidden variables depends on both settings.
Bart speculates only briefly about a physical basis
for his model. Even though his model does not accept all assumptions by \citet{PhysRev.47.777}, it proposes an answer to the title \textit{Can Quantum-Mechanical Description of Physical Reality Be Considered Complete?} of their paper: \textit{Not yet}.

The three positions are not reconciled, and we do not pretend otherwise.
Section \ref{discussion} is a discussion in which each author,
again in the first person,
states his opinions of the other two's proposals -- where he agrees and where he is unconvinced.
Section \ref{conclusions} records what we all agree on and notes where we differ.
Thus, we are offering three ways of finding comfort with Bell's proof and the experiments that have confirmed its conclusion.

The paper is written by three authors with overlapping but distinct backgrounds -- mathematical statistics and the theory of causality,
mathematical statistics and the foundations of quantum mechanics,
and a study of physics followed by a career in language technology, while keeping an eye on the never-ending debate between Bohr and Einstein about the philosophical status of quantum mechanics.
We have kept the first-person voice in Sections \ref{RiHeBa} and \ref{discussion} 
because the topic is one on which the authors of a joint paper on Bell's theorem are,
in our experience, entitled to disagree openly,
and because we think the reader is better served by three different positions than by an artificial compromise.

\section{Half of Bell's theorem: the classical side}\label{Ch2}

In Bell's later works he introduced a concept which he called \emph{local causality}. It corresponds to Einstein's \citep{EINSTEIN1936349} insistence that the quantum mechanical concept of joint state of a compound system cannot actually represent the actual states of the components. Einstein has no need at all of EPR's concept of elements of reality; he simply insists that the true states of two subsystems can only be correlated through their past interaction. Quantum mechanics (if correct; he sees no reason to doubt that) is incomplete because it does not include the physical variables which are responsible for the correlation. The separation of two particles created together at a source would lead to their joint quantum state, on separation, being a statistical mixture of product states.

\begin{figure}
\centering
\begin{subfigure}{.5\textwidth}
  \centering
  \includegraphics[width=.9\linewidth]{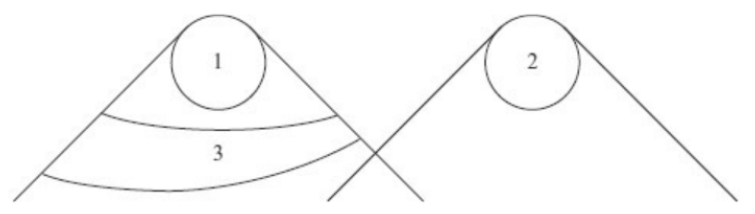}
  \caption{Full specification of what happens\\in 3 makes events in 2 irrelevant for predictions \\about 1 in a locally causal theory}
  \label{fig:bell4}
\end{subfigure}%
\begin{subfigure}{.5\textwidth}
  \centering
  \includegraphics[width=.9\linewidth]{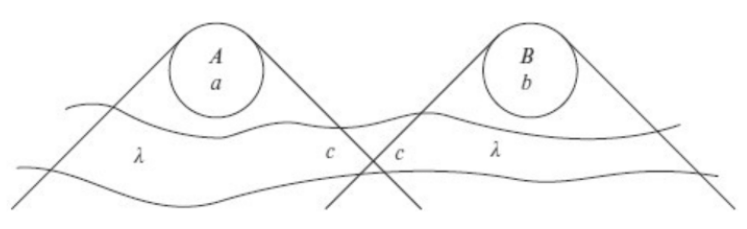}
  \caption{Diagram for CHHS inequality derivation}
  \label{fig:bell6}
\end{subfigure}
\caption{Subfigures (a) and (b) are reproductions of Bell's (2004) \emph{Nouvelle Cuisine} Figures 4 and 6, with Bell's captions. In our subfigure (a), the regions 1, 2 and 3 are regions in space-time, showing the past light-cones of regions 1 and 2. In subfigure (b), the variable $c$ stands for variables already known and accounted for, $\lambda$ for the supplementary hidden variables not accounted for in standard quantum mechanical modelling. $A$ and $B$ are the outcomes, $a$ and $b$ are the settings. [Bell's CHHS is of course nowadays known as CHSH]}
\label{fig:bell46}
\end{figure}

Bell explained his ideas very lucidly in his last paper, entitled ``La Nouvelle Cuisine'', \citet{Bell_Aspect_2004_24}, which included the following two illustrations, Figure \ref{fig:bell46}. 

\begin{figure}[H]
\noindent\centerline{\includegraphics[width=0.95\textwidth]{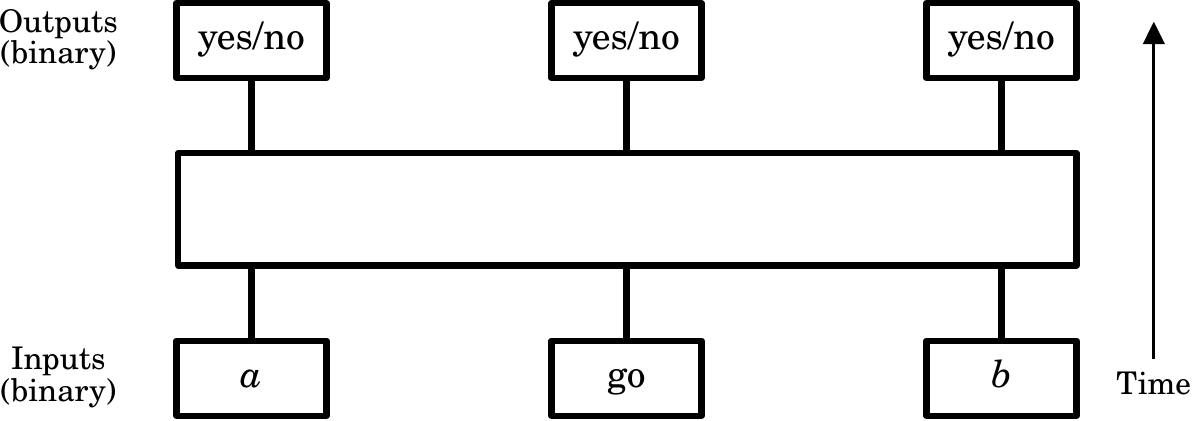}}
\caption{Spatio-temporal disposition of one trial of a Bell experiment.
Distance (left to right) is so large that a signal travelling from one side to the other at the speed
of light takes longer than the time interval between input and output on each side.
One ``go = yes'' trial has binary inputs and outputs; modelled as random variables $A, B, X, Y$.
\small (Figure 7 from J.S. Bell (1980), ``Bertlmann's socks and the nature of reality'')}
\label{fig1}
\end{figure}

\begin{figure}[H]
\noindent\centerline{\includegraphics[width=0.95\textwidth]{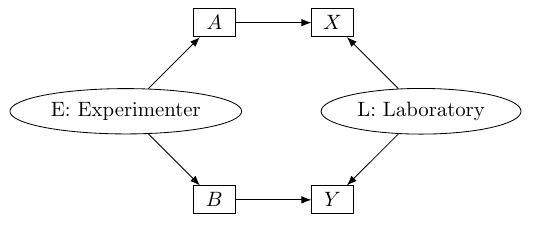}}
\caption{Graphical model of one trial of a Bell experiment}
\label{fig2}
\end{figure}

The reader of this Section will need some acquaintance with the theory of causality pioneered and promoted by Judea Pearl in his book \emph{Causality} \citep{Neuberg_2003}. We will study the causal model described in the DAG (directed acyclic graph) of Figure \ref{fig2}, which represents a classical physical view of part of what goes on in a Bell experiment.
Figure \ref{fig1} gives us some physical background and motivation for the causal model described in Figure \ref{fig2}. And the two figures in Figure \ref{fig:bell46} show the locations in space-time of both hidden and non-hidden variables which determine the probabilities of what happens in the two measurement stations.

How a Bell-type experiment is engineered in a quantum physics lab, and it can be engineered in several quite different ways, will not be explained here. But as well as the traditional Alice and Bob there can be a critical role for a third player Carol, at the intermediate location in Figure \ref{fig1}. These three ``persons'' are fictional, the whole process will be completely automated. During an experiment many trials are done. Again and again, Alice, Bob each take a binary action (press one of two buttons, flip a switch, \dots) and record a binary response (for instance, a light goes on or off). Carol has no choice of input but does take an action and soon thereafter observes a binary response. The actions of the three players follow a pre-chosen time schedule and are synchronised in the sense that the three \emph{outputs} are recorded at the three locations so quickly, that the speed of light should prevent distant inputs playing any role in the physical process going on between input to output at the other locations. For instance, Alice and Bob's inputs have no time to influence Carol's output.

The whole experiment thus generates a sequence of values of two binary inputs and three binary outputs, each set belonging to one ``time slot''. The aim of the experiment is to study the correlations between Alice and Bob's outcomes, conditional on their inputs and conditional on Carol's output. The statistical analysis of the data will therefore be restricted to the subset of time-slots in which Carol's outcome suggested that quantum interaction had a chance to take place between the physical systems in Alice and Bob's labs. This is called a Bell experiment with event-ready detectors, and the causal model represented by the DAG in Figure \ref{fig2} is a causal model for just one trial in which Carol got the outcome ``yes''.

Each trial accepted by Carol has two binary inputs or settings, and two binary outputs or outcomes. We will denote them using the language of classical probability theory by random variables $A, B, X, Y$ where $A,B\in\{1, 2\}$ and $X, Y\in\{-1, +1\}$. The settings $A$, $B$ should be thought of merely as labels (categorical variables); the outcomes $X,Y$ will be thought of as numerical. 
We will derive inequalities for the four theoretical expectation values, often called \emph{correlations} $\mathbb E_{ab}(XY) := \mathbb E(XY | A=a, B= b)$.
In Figure \ref{fig2}, the nodes labelled $A$, $B$, $X$, $Y$ correspond to the four observed binary variables in one trial accepted by Carol's output. The other two nodes annotated ``E: Experimenter'' and ``L: Laboratory'' correspond to factors leading to the statistical dependence structure of $(A, B, X, Y)$ of two kinds. On the one hand, the experimenter externally has control of the choice of the settings. In some experiments they are intended to be the results of external, fair coin tosses. Thus the experimenter might try to achieve that $A$ and $B$ are statistically independent and completely random. ``Experimenter'' and ``Laboratory'' are root nodes, they represent statistically independent variables (or groups of variables). The model assumes that the process leading to selection of two settings is statistically independent of whatever randomness goes on inside the long horizontal box of Figure \ref{fig1}, conditional on suitable features in the past, see Speakable and Unspeakable \citep{Bell_Aspect_2004_24} chapter 24 section 9 and especially Figure 6. That mechanism is unknown and unspecified, but of classical nature. Any apparent randomness is due to uncontrolled statistical variation in initial conditions. In the physics literature, one uses the phrase ``hidden variables''. The model represents a classical physical model, classical in the sense of pre-quantum theory, in which experimental settings can be chosen in a statistically independent manner from the uncontrolled parameters of the physical processes, essentially deterministic, which together with the setting at each end lead to the actually observed measurement outcomes.

Thus we are making the following assumptions on the probability distribution of the four binary variables observed in one trial accepted by Carol. There are two statistically independent random variables (not necessarily real-valued), which we will denote by $\lambda_E$ and $\lambda_L$, such that the probability distribution of $(A, B, X, Y)$ can be simulated as follows. First of all, draw outcomes $\lambda_E$ and $\lambda_L$, independently, from any two probability distributions over any two measure spaces. Given $\lambda_E$, draw outcomes $a, b$ from any two probability distributions on $\{1, 2\}$, depending on $\lambda_E$. Then, given $a$ and $\lambda_L$, draw $x\in \{-1, +1\}$ from some probability distribution depending on those two parameters. Similarly, independently, draw $y \in\{-1, +1\}$ from some probability distribution depending on $b$ and $\lambda_L$.

Thus, $\lambda_L$ is the hidden variable responsible for possible statistical dependence between $X$ and $Y$ given $A$ and $B$. 

In the theory of graphical models, one knows that such models can be thought of as deterministic models, where the random variable connected to any node in the DAG is a deterministic function of the variables associated with nodes with direct directed links to that node, together with some independent random variable associated with that node. In particular therefore, in obvious notation,
$$X = \phi(A, \lambda_L, \lambda_X),$$
$$Y = \gamma(B, \lambda_L, \lambda_Y),$$
where $\lambda := (\lambda_L, \lambda_X, \lambda_Y)$ is statistically independent of $(A, B)$, the three components of $\lambda$ are mutually independent of one another, and $\phi$, $\gamma$ are some functions. We can now define functions $f$ and $g$ and rewrite the last two displayed equations as
$$X = f(A, \lambda),$$
$$Y = g(B, \lambda),$$
where $f$, $g$ are some functions and $(A, B)$ is statistically independent of $\lambda$. This is what Bell called a local hidden variables model. 

So, thanks to the assumption that $(A, B)$ is statistically independent of $\lambda$, one can mathematically construct four random variables $(X_1, X_2, Y_1, Y_2)$ as
$$X_a = f(a, \lambda)$$
$$Y_b= g(b, \lambda).$$
These four have a joint probability distribution by construction, and take values in $\{-1, +1\}$, and are statistically independent of $(A, B)$. By the usual simple proof, see below, the Bell-CHSH inequalities hold for the four correlations $\mathbb E(X_a Y_b)$. But, by the statistical independence assumption, each of these four correlations is identically equal to the ``experimentally accessible'' correlations; defined as follows: for all $a, b$
\begin{equation}
	\begin{aligned}\label{2.1}
\rho_{ab} := \mathbb E(XY \mid A=a, B = b).
	\end{aligned}
\end{equation}
This equation looks obvious, but actually it needs a careful proof and the proof uses the independence assumption, since otherwise, conditioning on $(A, B)$ changes the distribution of what we can now think of as \emph{the} ``hidden variable'', the four-tuple $(X_1, X_2, Y_1, Y_2)$.

Finally, as we will show at the end of this Section,
\begin{equation}
	\begin{aligned}\label{2.2}
		-2 ~ \le ~ \mathbb E(X_1Y_1) - \mathbb E(X_1Y_2) - \mathbb E(X_2Y_1) - \mathbb E(X_2Y_2) ~ \le ~ +2 
	\end{aligned}
\end{equation}
and therefore
\begin{equation}
	\begin{aligned}\label{2.3}
		-2 \le \rho_{11} - \rho_{12} - \rho_{21}-\rho_{22} \le 2 
	\end{aligned}
\end{equation}
and similarly for the comparison of each of the other three correlations with the sum of the others. Equation \eqref{2.3} combines two separate inequalities. Each one is called a one-sided Bell-CHSH inequality; the combination \eqref{2.3} is called a two-sided Bell-CHSH inequality. 

So far, we have argued that the physical assumptions of \emph{local realism} together with the statistical independence assumption lead to a mathematical model for one trial in a Bell-CHSH type experiment where Alice and Bob each toss a coin to select a measurement setting on a measurement apparatus, and then go on to observe a binary outcome, of the following form. The outcomes of the two coin tosses are denoted by $A$ and $B$, they take the values in the set $\{1, 2\}$. These are just labels. The observed measurement outcomes will be denoted by $X$, $Y$, they take values in $\{-1, +1\}$. In one \emph{trial} one observes one four-tuple $(A, B, X, Y)$. According to local realism, the probability distribution of the observed random variables has the following structure. Mathematically, the classical probability space on which $(A, B, X, Y)$ is defined can be extended to include four ``counterfactual outcomes'' $(X_1, X_2, Y_1, Y_2)$, such that $X= X_A$ and $Y= Y_B$ and such that the pair $(A, B)$ is statistically independent of the four-tuple $(X_1, X_2, Y_1, Y_2)$. According to this probability model, the coin tosses \emph{selected} which of the counterfactual outcomes $X_i$ and $Y_j$ got observed. The coin tosses need not be fair and need not be independent of one another. One can imagine a pair of coins being manufactured in advance and having correlated bias. But this physical process should be independent of the physical processes going on inside the measurement stations. The coin toss outcomes could even be fixed in advance, there is no need to suppose they are random.

We already took account of the assumption of \emph{locality} by giving $X_i$ and $Y_j$ each just one index. The outcome which Alice would have seen had she chosen setting $i$ and Bob chosen setting $j$, which one could denote by $X_{ij}$, is such that $X_{i1} = X_{i2} =: X_i$. We are not assuming that it physically exists, whatever that should mean (a metaphysical question, thought by Einstein to be irrelevant). It is made plausible in experiments by the spatio-temporal constraints concerning insertion of settings or inputs (one in each wing of the experiments) and recording of outcomes or outputs (also one in each wing of the experiment). Bob's outcome must have been registered before Alice's setting choice could have influenced its realisation at Bob's location, even if information was travelling at the speed of light.

The assumption $X = X_A$ is sometimes called \emph{counterfactual definiteness}. The notation uses standard probabilist's shorthand, it actually stands for $X(\omega) = X_{A(\omega)}(\omega)$ for all {$\omega\in\Omega$}, an underlying probability space on which all these random variables are defined. Again, it is not a metaphysical assumption, an assumption about reality. It is an assumption within a mathematical model, a statistical model for the probability distribution of $(A, B, X, Y)$ observed in a single trial.

The assumption of \emph{local realism} is the assumption that $(X_1, X_2, Y_1, Y_2)$ can be (mathematically) defined at all. A local-realistic mathematical model of the physics going on in this experiment allows one to include in the model, in a consistent way, what the outcomes would have been, had certain physical settings been different from what they were in actuality, and so that Alice's potential outcomes don't depend on Bob's actually chosen setting.

The crucial assumption of \emph{statistical independence} between settings and counterfactuals is an assumption of \emph{freedom} or of \emph{no-conspiracy}. 

The three assumptions (locality, realism, no-conspiracy) are mathematical assumptions concerning the mathematical existence of a probabilistic model which reproduces certain predictions of quantum mechanics. The words have a long history and are associated with philosophical positions and past controversies concerning the philosophy of science. Our terminology is nowadays pretty standard. Moreover, {``counterfactual definiteness''} has become a standard term in the modern scientific and statistical understanding of \emph{causality}. Counterfactual reasoning is common currency in medical statistics, in epidemiology. It is essential in moral and in legal reasoning, and in the understanding of (and learning from) history. Statisticians say that they can only establish correlations, not causation, but they are hired to establish causation. 

A further side remark is that Bell, following EPR \citep{PhysRev.47.777}, originally derived realism from locality, by noting that with equal settings, outcomes were equal and opposite. It is hard to conceive that this could be the case if the outcome on each side of the experiment was not actually predetermined in some way or other. This leads furthermore to the idea that all randomness in the measurement outcomes is purely due {to} randomness at the source. However, for very sound reasons (in experiments one does not observe \emph{perfect} anti-correlation at equal settings; at best, only approximate anti-correlation) Bell and the whole community rapidly adopted the CHSH inequality and allowed for further randomness at the measurement locations.

Given these assumptions, let us take a look at the following expression $Z:= X_1 Y_1 - X_1 Y_2 - X_2 Y_1 - X_2 Y_2$. One can rewrite it as $X_1(Y_1 - Y_2) - X_2(Y_1 + Y_2)$. The four random variables $X_i$ and $Y_j$ take the values $\pm 1$, so either $Y_1 = Y_2$ or $Y_1 = - Y_2$, so one of the two terms in brackets equals zero, the other equals $\pm 2$. They are each multiplied by $\pm 1$ so the value of the whole expression is $\pm 2$. That implies that its expectation value lies between $-2$ and $+2$. Therefore, by linearity of the expectation operation, 
\begin{equation}
	\begin{aligned}\label{2.4}
		-2 ~\le~ \mathbb E(X_1 Y_1) - \mathbb E(X_1 Y_2) - \mathbb E(X_2 Y_1) - \mathbb E(X_2 Y_2) ~\le~ 2.
	\end{aligned}
\end{equation}
By statistical independence,
\begin{equation}
	\begin{aligned}\label{2.5}
		\mathbb E_{ij} (XY) := \mathbb E(X Y \mid A = i, B = j) = \mathbb E(X_i Y_j \mid A = i, B = j) = \mathbb E(X_iY_j ).
	\end{aligned}
\end{equation}
This leads to the following constraint on four experimentally accessible quantities, a two-sided Bell-CHSH inequality:
\begin{equation}
	\begin{aligned}\label{2.6}
		 -2 ~ \le ~ \mathbb E_{11}(X Y) - \mathbb E_{12}(X Y) - \mathbb E_{21}(X Y) - \mathbb E_{22}(X Y) ~ \le ~ 2.
	\end{aligned}
\end{equation}

This is one of the complete set of 4 two-sided Bell-CHSH inequalities. \citet{PhysRevLett.48.291} has shown that in the presence of the no-signalling equalities, namely that the conditional distribution of $X$ given $A$ and $B$ does not depend on $B$ and the conditional distribution of $Y$ given $A$ and $B$ does not depend on $A$, the complete set of CHSH inequalities is equivalent to the mathematical existence of the hidden variables model just described, the ``local hidden variable'' being $(X_1, X_2, Y_1, Y_2)$, independent of $(A,B)$. Note that it is not the hidden variable that is local, it can include components from all over the experiment. As Bell emphasised, the adjective ``local'' applies to how the \emph{settings} $A$ and $B$ influence the outcomes $X$ and $Y$. Various other inequalities \citep{Clauser2017,PhysRevD.10.526,PhysRevA.47.R747} are equivalent to Bell-CHSH inequalities under the same no-signalling condition. Bell's original three correlation inequality \citep{PhysicsPhysiqueFizika.1.195} is equivalent to CHSH too when one of the four correlations equals +/-1.

In summary, we have used the assumption of local causality to derive counterfactual measurement outcomes as functions of classical hidden variables responsible for the correlations between measurement outcomes given measurement settings, and such that the distribution of those hidden variables does not depend on the settings.
From this we derived the Bell-CHSH inequalities for the experimentally determined correlations between the factual measurement outcomes, given each pair of factually chosen measurement settings.

The material in this Section is taken from two publications \citep{Gill_2014,gill2023bellstheoremexercisestatistical}.

\section{The Bell experiments}\label{Ch3}

\subsection{Loopholes in Bell tests}

Bell experiments are designed to test the local reality assumption originally formulated by \citet{PhysRev.47.777} (EPR) and used by \citet{PhysicsPhysiqueFizika.1.195} and \citet{PhysRevLett.23.880} in the derivation of what we call the `CHSH inequality'. 

The derivation of the CHSH inequality features several types of events in Bell experiments that are assumed to not causally affect other events: the detection of entanglement, the decision to choose a particular setting, the realisation of a chosen setting, the click in a detector announcing a particular outcome of a spin component measurement.

In a loophole-free Bell test, care has been taken to ensure that all assumptions needed to derive the inequality are met.
It is conceivable that a test violates the CHSH inequality for a reason other than the validity of QM if not all locality conditions are met. Such a test has a loophole.
If we cannot assert that every assumption has been addressed in the experimental setup, the acceptance of the outcome of the experiment is conditioned on our belief that the loopholes have no effect. In the words of \citet{Larsson_2014}, a loophole `refers to circumstances in an experiment that force us to make extra assumptions for the test to apply'.

Larsson discusses the following loopholes:

\subsubsection{The locality loophole}
The locality assumption states that measuring one particle
could not have any influence on another, distant, particle \citep{gill2023optimalstatisticalanalysesbell}.
The direct causal isolation of these four events can be ensured by spacelike separating Alice's and Bob's setting decisions and measurement outcomes.

\subsubsection{Freedom of choice loophole}
It is assumed that a choice of setting by Alice or Bob has no cause in common with any other event in a Bell experiment, especially not the system under observation. This is understood by some experimental groups to mean that the choice of a setting is not in the future light cone of the detection of entanglement. This can be realised in several ways. The most radical way is to let choices be decided by events observed at opposite sides of the visible universe. A less radical way is to create an `event ready' setup, in which the detection of entanglement is not the earliest event in a trial, but postponed somewhat to make sure that the future light cone of the detection event does not encompass the choices of settings.

\subsubsection{The fair sampling or detection loophole}
Bell experiments, especially those that use photons, lose many trials because one or both detections do not happen. Fair sampling means that the proportion of detected versus undetected events of a particular combination of outcomes is the same for all combinations of outcomes. The experiment may be skewed towards the success of pairs of measurements that build up to the violation of the CHSH inequality if fair sampling cannot be ascertained.

In a Bell experiment that uses single-channel analysers, the efficiency of the detectors is important. If the efficiency of the detectors is low, the maximum of the CHSH expression is low as well, to the point that the CHSH inequality cannot be violated. Depending on some other assumptions, the minimum detector efficiency needed to violate the CHSH inequality is 82.84\% to 85.97\%. \citet{PhysRevA.47.R747} showed that with non-maximally entangled states (Eq. \eqref{Eber}), the minimum detector efficiency needed for violation of the CHSH inequality is much lower, 66.67\% when $r\rightarrow0$. However, the noise tolerance becomes lower as the efficiency approaches this low limit.
\begin{equation}
	\begin{aligned}\label{Eber}
		\psi=\frac{\ket{\leftrightarrow \updownarrow}+r\ket{\updownarrow \leftrightarrow}}{\sqrt{1+r^2}}
	\end{aligned}
\end{equation}
Larsson reports that still lower bounds are possible, but impractical.

\subsubsection{The memory loophole}
The last independence assumption that has to be taken care of is that a trial in a Bell experiment must not be influenced by the history of previous trials. For example, if the choices of settings follow a regular temporal pattern, a memory of earlier settings could conceivably influence the outcome of the current trial.

The memory loophole is handled by making sure that the settings in each trial are chosen at random, so that no temporal pattern arises. A more radical way to close this loophole is to distribute a large number of copies of the experimental setup over space, and to perform only a single trial in each setup in such a way that each trial is completely outside the future light cones of all other trials. In addition, the memory loophole should be handled by using martingale statistics. This also fixes the problem of drifts and shifts in the physics as time progresses. \citep{Gill_2014}

\subsubsection{The coincidence loophole}
Especially in experiments where trials follow each other at a high rate, one has to decide on a protocol that decides whether an observation at A and an observation at B belong to the same trial. This can be done using a coincidence window: if and only if two observations happen in a time frame no longer than the coincidence window, it is safe to assume that they belong to the same trial. The coincidence assumption is that this protocol is blind to the actual outcomes. If, on the contrary, the protocol is more critical of observation pairs of a particular signature than observation pairs with other signatures, then the Bell experiment will give a skewed estimate of the CHSH expression.

\subsubsection{The finite statistics loophole}
If the number of observations is low a local-realist model can produce an apparent violation by chance.

\subsubsection{The loophole of low violation}
Authors often report the amount of violation of the CHSH inequality as a number of sample standard deviations. The underlying but unwarranted assumption is that the distribution of the observations is known but cf. \citet{a7d20790ac20477b94cb1d80c1693fdb,Elkouss2016} for guaranteed p-values on the condition that the settings are chosen again and again completely at random

\subsubsection{The postselection loophole}
Larsson discusses a loophole that perhaps is better called a flawed design. He calls it the ``postselection loophole''. This loophole refers to the fact that in the Franson interferometer Bell test \citep{PhysRevLett.62.2205}, only 32 of all 64 combinations of settings, outcomes, and photon paths are taken into account. This makes it possible to design a LHV theory that violates the CHSH inequality, see \citet{PhysRevLett.83.2872}. \citet{PhysRevLett.102.040401} show how the design of the interferometer has to be changed to eliminate this loophole.

\subsection{Aspect's Bell tests}
In 1981 and 1982, Alain Aspect led three Bell tests. In all three tests the separation between the `Alice' and `Bob' setups was 12-13~m and in all three tests, measurements were performed on correlated photon pairs at wavelengths 422.7~nm and 551.3~nm obtained by two-photon excitation of a $\left(J=0\right)\rightarrow\left(J=1\right)\rightarrow\left(J=0\right)$ cascade in $\leftidx{^{40}}{\text {Ca}}{}$. The first test \citep{PhysRevLett.47.460} did not test the CHSH inequality. In the second test \citet{PhysRevLett.49.91}, each of the two measurement sites consisted of a rotatable two-channel polariser that would send a photon one or the other way, depending on the linear polarisation of the photon. Photomultiplier tubes (PMT) were placed in each of the two optical paths to detect the photons after they had passed the polariser. Due to low efficiencies, not all trials resulted in a successful pair of measurements. Only if two measurements were reported in a coincidence time window of 20 ns, corresponding to 6 m as the photon flies, it was assumed that the trial had succeeded. The experiment indicated that the CHSH inequality was violated by a wide margin, $S_{exp} =2.697\pm 0.015$.
Alice's polarimeters were manually set to settings $a_1$ or $a_2$, and Bob's to $b_1$ or $b_2$, respectively. The extended periods in which the settings remained unchanged made it in principle possible for Alice's setting to be shared with Bob and to influence Bob's measurement, and vice versa. There was no guarantee that the static settings did not exert an influence on the outcomes of the other party. Aspect's first test of the CHSH inequality was therefore susceptible to the locality loophole.

In their third published test, \citet{PhysRevLett.49.1804} mitigated the locality loophole by very quickly switching the light paths in Alice and Bob's apparatuses towards one of two polarisers, thereby doubling the number of polarisers. Alice's two polarisers were preset to directions $a_1$ and $a_2$, while Bob's were preset to directions $b_1$ and $b_2$. The switching was done periodically and at a frequency so high that the actual setting during a polarisation component measurement could not be signalled to the other site in time to influence the outcome at the other site. Although the locality loophole was circumvented, the memory loophole remained, because the settings came in a regular temporal pattern.

With the third experiment, Aspect set the trend for later Bell experiments that tested entangled photons. Even though each measurement site had to handle not one but two settings, Aspect did not double the number of PMTs from four to eight. This had the consequence that only one of the two possible outcomes, e.g. `+', resulted in a detection. Since the experimental verification of the expectation values predicted by QM would require that both possible outcomes, `-' and `+', were detected, Aspect used the Clauser-Horne inequality, which does not require expectation values but probabilities:
\begin{equation}
	\begin{aligned}\label{S++ineq}
		-1 \le S_{++} \le 0
	\end{aligned}
\end{equation}
where 
\begin{equation}
	\begin{aligned}\label{S++}
		S_{++} = p_{++}\left(a,b\right)-p_{++}\left(a,b'\right)+p_{++}\left(a',b\right)+p_{++}\left(a',b'\right)-p_{+}\left(a'\right)-p_{+}\left(b\right)
	\end{aligned}
\end{equation}
In this expression, $p_{++}\left(a,b\right)$ is the probability that both PMTs detect a photon while Alice's setting is $a$ and Bob's setting is $b$, and \textit{mutatis mutandis} if the settings $a'$ and $b'$ apply. The probability $p_{+}\left(a'\right)$ is the probability that Alice's PMT detects a photon while the setting is $a'$, irrespective of whether Bob's PMT detects a photon or not, while $p_{+}\left(b\right)$ is the probability that Bob's PMT detects a photon while the actual setting is $b$.

QM violates Eq. \eqref{S++ineq} maximally if the measured state is maximally entangled ($P\left(+\right) = P\left(-\right)$). For photon entanglement $p_{++}\left(\alpha\right) = p_{--}\left(\alpha\right) = \frac{1+\cos(2\alpha)}{4}$ and $p_{+-}\left(\alpha\right) = p_{-+}\left(\alpha\right) = \frac{1-\cos(2\alpha)}{4}$. Let $a=0^\circ$, $a'=45^\circ$, $b=22.5^\circ$ and $b'=67.5^\circ$. Then the value $S_{++}$ is well above the upper limit in Eq. \eqref{S++ineq}:
\begin{equation}
	\begin{aligned}\label{S++viol}
		S_{++} =& p_{++}(22.5^\circ)-p_{++}(67.5^\circ)+p_{++}(22.5^\circ)+p_{++}(22.5^\circ)-\frac{1}{2}-\frac{1}{2}\\
		=&\frac{{1+\frac{1}{\sqrt{2}}}}{4}-\frac{{1-\frac{1}{\sqrt{2}}}}{4}+\frac{{1+\frac{1}{\sqrt{2}}}}{4}+\frac{{1+\frac{1}{\sqrt{2}}}}{4}-1\\
		=&\frac{\sqrt{2}-1}{2}
	\end{aligned}
\end{equation}
It is easy to show that 
\begin{equation}
	\begin{aligned}\label{SasSik}
		S = S_{++} + S_{--} -S_{+-} -S_{-+}
	\end{aligned}
\end{equation}
where 
\begin{equation}
	\begin{aligned}\label{S++2}
		S_{xy} = p_{xy}\left(a,b\right)-p_{xy}\left(a,b'\right)+p_{xy}\left(a',b\right)+p_{xy}\left(a',b'\right)-p_{x}\left(a'\right)-p_{y}\left(b\right)
	\end{aligned}
\end{equation}
and $x \in \setdef{+,-}$ and $y \in \setdef{+,-}$.

\subsection{2015--2017: the first (almost) loophole-free Bell tests}
Enabled by progress in nano-scale and fibre optics engineering, 2015 saw three papers on successful loophole-free Bell tests. In 2017, a fourth article appeared.

Two of the experiments were event-ready tests, which do not rest on a fair sampling assumption. Therefore the tests were free of the fair sampling (or detection) loophole. Both these tests employed entangled spins of matter particles and tested the degree of violation of the CHSH inequality Eq. \eqref{2.6}. In October 2015, a group in Delft, the Netherlands, published the results of a Bell test on entangled solid-state quantum registers, namely nitrogen-vacancy (NV) defect centres in diamond \citep{Hensen2015}. The two sites were separated by 1280 m. The group reported a significance level (p-value) of 0.039. To circumvent the freedom of choice loophole, the experiment had a remarkable design. Usually, we depict a Bell test as an experiment in which entangled particles are separated and fly in opposite directions, to be measured when they are far apart. In the Delft experiment the events occurred in a different order. Entanglement was realised when photons entangled with the widely separated NV centres arrived at a device spatially between the two measurement sites. Once there, the photons were either found to be entangled, in which case the trial's spin component measurements were valid, or they were found to be not entangled, in which case the results of the trial's spin component measurements were discarded. The advantage of this order of events, which is called `entanglement swapping', is that the heralding of the entanglement happened at a point in space-time that was spacelike separated from both the choice of settings and the ensuing spin component measurements. In that way, the Delft group states, the freedom of choice loophole is eliminated.

In December 2015, a team in Vienna, Austria \citep{Giustina_2015} and a team at NIST in Boulder, Colorado \citep{Shalm_2015} simultaneously published the results of Bell tests on entangled photons. The results of both groups were highly significant, if expressed as p-values. These photon-based tests used a different analysis than the matter-based experiments. Instead of the better-known CHSH inequality, they tested the degree of violation of Eq. \eqref{S++ineq}. There are two main differences between matter and photon Bell experiments. Matter-based entanglement is harder to create than photon-based entanglement and happens at a much lower rate. On the other hand, photons are much harder to guide through the experiment. Many photons are lost in transit. Given the lower efficiency of photon trials, combined with the very high number of successful trials, the use of the CH inequality in the form proposed by \citet{PhysRevA.47.R747}, also known as the CH-Eberhard or CH-E inequality, was the only option for the photon teams \citep{Kofler_2016}. 

To close the memory loophole, the Vienna team did not assume independent identically distributed (IID) statistics. Instead, they used a statistical analysis based on a martingale model. In a martingale model the drift cannot become positive, so any experimentally established value greater than zero indicated a violation of the CH-E inequality \citep{gill2002accardicontrabellcum}. Using that model, they answered the question ``How likely is it that the data produced by the experiment resulted from a statistical fluctuation away from the baseline value that one would predict if local realism were true?''. In a Bell test, the p-value is the probability, assuming local realism, of observing a violation of a Bell-type inequality at least as large as the one observed. A very low p-value means that such a violation would be very unlikely if local realism were true, which is strong evidence against local realism. 

The Vienna group reported a significance level of $3.74 \cdot 10^{-31}$, while the NIST group reported a significance level of $2.3 \cdot 10^{-7}$. In Vienna the distance between the measuring sites was 58 m, in Boulder the distance was 184.9 m.

While the Vienna experiment had been underway for several years, the NIST group started later and had the advantage of newer and better components. The fibre optics were less critical in the NIST experiment because it employed frequencies in the telecom range for which there are very efficient fibres. The Vienna group, on the other hand, used photons with higher frequencies, which have higher energies and therefore are easier to detect. The NIST team used more optical fibre than the Vienna group. Due to the stress-induced birefringence of optical fibre, which changed with temperature variations, the whole setup had to be re-calibrated every 10 minutes. In that way, input and output polarisations were kept identical to within $99.9\%$.

The fourth test, which used entangled atoms, was published by a group in Munich, Germany, in 2017 \citep{Rosenfeld_2017}. They employed entangled $\leftidx{^{87}}{\text {Rb}}{}$ atoms that were separated by 398 m.
The Munich group attained a much better significance level than the Delft group, $2.57 \cdot 10^{-9}$. However, this p-value is an underestimate. A more accurate value is the much worse $7\permil$ \citep{gill2023optimalstatisticalanalysesbell}.
The space-time map of a trial in the Munich experiment was asymmetric. To establish a Bell state of two entangled Rb atoms, the Rb atoms were excited, whereafter the polarisations of photons emitted by spontaneous decay were entangled with the Rb atoms. The photon from site 2 was sent towards site 1, where it met the photon from site 1 in an interferometric Bell state measurement (BSM) setup, located close to site 1. When the BSM detected a two-photon coincidence in a particular detector combination, the two atoms were entangled and heralding signals were sent to site 1 and the far away site 2. Both sites waited just long enough to make sure that both sites `knew' that the atoms were entangled. Then settings were chosen at random and measurement results obtained within the short time span wherein the sites were spacelike separated, using the polarisations of readout lasers. In contrast to the Delft group, the Munich group did not attempt to herald the success of the entanglement at a space-time point that was spacelike separated from choosing the settings and measuring the spin components. The Delft group had these events spacelike separated to close the freedom of choice loophole. The Munich group does not claim to have closed the freedom of choice loophole. The Delft group argued that the hidden variable arises at the space-time point of the heralding and that, in their experiment, the HV could not influence the choice of setting. The Munich group disagrees. They deny that there is a particular moment at which hidden variables are defined in an event-ready test using entanglement swapping, in contrast to experiments with photon pairs. To close the freedom of choice loophole, they say, one would have to produce random numbers in a way that obviates any interaction with all other events of the Bell test. The NIST group also says that, in principle, the freedom-of-choice loophole cannot be closed, but they add that spacelike separation cuts off many ways in which this loophole could be used in a HV model.

\section{Some relevant literature}\label{Ch4}

There are very many recent articles referring to the Bell theorem and the results of the Bell experiment. In this paper, we will only refer to a few of these articles.

First, a side remark. It is a common belief that the Bell theorem is only concerned with the microworld. Arguments against this belief are given by \citet{Aerts2000,Aerts2006}, who in particular have studied the Bell inequalities in macroscopic systems, cognition and human language. 

The 2022 Nobel prize in physics was given to Alain Aspect, John F. Clauser, and Anton Zeilinger ``for experiments with entangled photons, establishing the violation of Bell inequalities and pioneering quantum information science''. But it is important to keep in mind that, as pointed out already by \citet{Mermin1981BringingHT} in a discussion of a similar thought experiment, the conclusions from such experiments are not only that quantum mechanics is an extraordinarily peculiar theory, but that the world is an extraordinarily peculiar place.

A surprisingly large number of authors have been so astonished by these conclusions that they have given various arguments against the Bell theorem in the form that has been explained in Section~\ref{Ch2} above. All these arguments may be countered; we give only one example here.

In a series of papers ending with \citet{Kupczynski_2020}, Marian Kupczynski gave arguments against the derivation of the CHSH inequality under certain conditions. He then states that `Nevertheless, imperfect correlations between clicks in spin polarisation correlation experiments may be explained in a locally causal way, if contextual setting-dependent parameters describing measuring instruments are correctly included in the description.' In particular, he argues that Bell's theorem can be circumvented if one takes correct account of contextual setting parameters describing measuring instruments.

\citet{gill2023kupczynskiscontextuallocallycausal} show that this is not true. Taking account of such parameters in the way that Kupczynski suggests, the CHSH inequality can still be derived. Thus quantum mechanics and local realism are not compatible with each other.

So, in the rest of this article, we will assume that Bell's theorem is valid: In order to understand the results of the recent Bell experiments, we cannot keep a world picture which is local, non-conspiratorial, and realistic. 

Some authors have concluded that the world must be non-local. They seem to claim that signals can pass instantaneously from Bob to Alice, which then is in contradiction to Einstein's relativity theory. Two examples of papers based upon such conclusions, are \citet{RevModPhys.86.419} and \citet{Maudlin_2014}. Brunner et al. define locality as the property that the responses of Alice and Bob, given their settings, are independent when conditioned upon some hidden variable. The review paper then addresses the technical problem on how one can show that the measurement statistics of a given experiment do not satisfy this condition. Maudlin explores consequences for Einstein's theory. He claims repeatedly that what Bell proved, was that the physical world is non-local, and that this non-locality is what bothered Albert Einstein, compare the famous EPR paper \citep{PhysRev.47.777}. He also in that connection says that the reality condition had better not apply if one is to maintain that the quantum-mechanical description is complete.

The example of the EPR paper, as modified by \citet{EARLE1951207} was already taken up in connection to Bell's theorem by \citet{Greenberger1989}, who showed that, using a slightly more complicated model than Bell, one also here cannot generally introduce deterministic, local models to explain the results.

\citet{palmer2024superdeterminismconspiracy} presents a locally causal, non-conspiratorial model defending superdeterminism, which violates Bell's ``Measurement Independence''. It proposes that entangled particles have unique properties, $\lambda$, linked to a discretised complex Hilbert space, where $\lambda$ fixes measurement settings $a$ and $b$.

\citet{Wiseman_2014} compares Bell's original theorem \citep{PhysicsPhysiqueFizika.1.195} with his newer theorem in \citet{Bell1976}, which states that quantum theory is incompatible with the single property of local causality. Although the conclusions of these two theorems are logically equivalent, their assumptions are not, and the different versions of the theorem suggest different conclusions, embraced by different communities. Wiseman's goal is to explore what can be done to increase mutual understanding.

\citet{BARR2024104134} explores Bell inequality violation in a completely new setting, high-energy physics, and shows that aspects and peculiarities of quantum physics are taking an increasingly central position in physics -- from quantum computers to information theory, from theoretical developments to innovative applications.

It is interesting that \citet{Fritz_2012,Fritz2016} recently went beyond Bell's theorem, and generalised his results to scenarios with arbitrary causal structure.

\section{Views by the authors}\label{RiHeBa}

\subsection{Richard Gill}\label{Richard}

In this Subsection, I will state where I come to after the argument of Section~\ref{Ch2} and the experimental record of Section \ref{Ch3}, together with certain theoretical developments 
which have not been covered yet: Belavkin's theory of ``eventum mechanics'' \citep{Belavkin1994}, 
about which I wrote a tutorial introduction \citep{gill2022belavkin}.

My background is in mathematical statistics, and I have long advocated that Bell's theorem is best read as an exercise in the statistical theory of causality in the tradition of \citet{Neuberg_2003}.
The directed acyclic graph of Figure \ref{fig2} is derived from a concept which Bell called local causality in \citet{Bell_Aspect_2004_24}. It encodes three assumptions: \emph{locality}, \emph{realism}, and \emph{no-conspiracy} (the statistical independence of settings from whatever hidden mechanism gives rise to outcomes).
Counterfactual definiteness in a mathematical sense and statistical independence, and hence the Bell-CHSH inequality, follow from local causality.
The experiments of Section \ref{Ch3} rule the conjunction of locality, realism and no-conspiracy out. One cannot hold on to all three at once. One is moreover forced to give up local causality.

The literature contains three broad escape routes from Bell's theorem, and I find each of them more costly than what they seek to avoid.

The first route is to weaken no-conspiracy, i.e.\ to accept superdeterminism or retrocausality. The physics of a coin toss, or of a conscious choice of setting, or of a state-of-the-art pseudo-random generator, would then be inseparable from the physics of the photodetector clicks at the other side of the lab. I agree here with \citet{Larsson_2014}: the loophole of superdeterminism cannot be closed by scientific methods; the assumption that the world is \emph{not} superdeterministic is needed to do science in the first place. Proposals in this direction \citep[e.g.][]{palmer2024superdeterminismconspiracy} can explain anything and therefore, in the Popperian sense, refute nothing. Accepting them means surrendering reductionism, and with it the explanatory ambitions that made physics what it is.

The second route is to weaken locality in the sense of admitting a genuine action-at-a-distance. This is the direction taken by, for example, \citet{Maudlin_2014} and in some readings of the review by \citet{RevModPhys.86.419}. I do not find it a stable resting place either. As I have argued in correspondence with my co-authors, realism and locality cannot be defined independently of one another: the less one takes to be real, the easier it becomes to maintain that what remains is local. A commitment to non-local \emph{influences} between outcomes still presupposes a robust ontology of outcomes and of the settings that supposedly influence them; and the no-signalling equalities forbid using any such influence to send a message. I prefer not to pay an ontological price whose only use is to protect an intuition that the experiments have already contradicted.

The third route is the route of the ``Bell deniers'' \citep[e.g.][]{Kupczynski_2020}, who seek a flaw in the derivation itself. Justo Pastor Lambare and I have examined the main recent attempts \citep{gill2023kupczynski,gill2023bellstheoremexercisestatistical} and shown that, properly formulated, the corresponding ``contextual'' or ``probabilistic coupling'' schemes do not evade the three assumptions of the theorem; the CHSH inequality is re-derived. There is no mathematical crack through which classical local realism can slip.

What remains is what I choose to give up, and it is the one that I find we can most honestly do without: the demand for a \emph{mechanistic} explanation of each single outcome. Bell's theorem, together with its experimental confirmations, tells us that there are events in the real world which are uncertain and which have well-defined probabilities, but for which there does not exist a local, classical mechanism that ``explains'' why one outcome occurred rather than another. To my mind this suggests that Aristotle was wrong: some events do not have causes in the sense he had in mind. Irreducible randomness is part of the basement-level fabric of reality, and determinism has its limits. And moreover, irreducible randomness can be ``non-local''.

It is worth stressing, in the language of \citet{Bell_Aspect_2004_book}, that none of this forces us to adopt a new ``quantum probability'' as a rival to the Kolmogorov theory under which classical statistics operates. In my essay \citep{gill2023bellstheoremexercisestatistical} I argue that the concept of probability is the same in both settings: a number between zero and one attached to an event or proposition, behaving under the usual calculus. What differs is the \emph{origin} of the randomness. Classical randomness can always, in principle, be reduced to uncontrolled variation in deterministic initial conditions. Quantum randomness, on the Bell-experimental evidence, cannot. Every classical probability space is a quantum probability space, but the converse fails: quantum theory has no universal set $\Omega$ on which all observables are simultaneously defined, and no dispersion-free states. That structural difference, and not a different meaning of the word ``probability'', is what Bell's theorem forces on us.

I am a statistician, and I take seriously George Box's aphorism that all models are wrong but some are useful. ``Physical reality'' is itself a model -- a shared, useful fiction without which the collective enterprise of science would not hold together. We do science together because we agree to play the game that a physical reality exists. Within that game, quantum mechanics is an exceptionally successful predictive model. The conclusion I draw from the Bell experiments is not that this game should be abandoned, but that we should resist demanding from it more than it can deliver: a picture in which every outcome has, in principle, a local classical cause. Quantum phenomena lie beyond that kind of understanding. My comfort with the Bell proof and the Bell experiments, to use the title of this paper, is therefore the comfort of someone who has learned to accept irreducible quantum randomness as simply part of how the world works.

Still, there are features of conventional quantum mechanics which remain for me cause for discomfort, and one of them is the Schr\"odinger cat problem. I like to see it as a paradox, because it is a problem which I believe can be resolved, using the concepts introduced by \citet{Belavkin1994} in numerous unfortunately hard to read papers. The problem is to resolve the apparent conflict between deterministic unitary evolution and randomness of outcomes of a measurement. ``Measurement'' is surely also a physical process and quantum mechanics should describe it. The evolution of a so-called quantum system interacting with a measurement device should be describable as the Schr\"odinger evolution of a bigger, joint system. But then, after measurement, we still have a quantum superposition of a live cat and a dead cat.

The conventional solution has been to adopt a so-called Heisenberg cut. Our model of the world has to have two levels: the top, macroscopic, level is classical, but underlying that is a microscopic quantum world. When applying quantum physics to the real world, one has to place the cut somewhere. 

Belavkin has a different picture: the cut is not between macroscopic and microscopic, but between past and future. In a slogan, ``the past is particles, the future is a wave''. 
But there is no ``observer'' in the theory. Past and future are always relative to a point in space-time, which I will refer to as ``here and now''. 
For utter simplicity I take time as discrete. Given are a Hilbert space, a quantum state on that space, and a unitary evolution.
Think of this as a model of a tiny toy discrete time universe.
Consider a set of observables which commute with one another and which are compatible with the unitary evolution of our toy universe in the sense that backward-in-time translations or versions of an observable in that set are also in that set. For instance, if ``position now'' (of some particle) is in the set, then one demands that it commutes with position of the same particle at each moment in the past or future. In that case, given the quantum state of the whole system, the whole past history of the particle's position at a particular time, ``now'', can be assigned a value, which is just its path, looking back into the past up to and including now. Its future locations also commute with one another. They can be assigned, using the Born rule, a probability distribution. These ingredients are compatible with one another, in the sense that we end up with a description of a stochastic process extending from as far back in time as we like, and as far forward in time as we like. The values assigned by the Born rule to all of the special set of commuting observables at a given time form a Markov process. The state space of the Markov process stands for ``everything you need to know now, in order to predict the future''. 

At this point, my overall conclusion is that I feel obliged to abandon Bell's \emph{local causality}. I would however be happy to add a qualifying adjective to that expression. Bell defines a concept which could best be called \emph{classical local causality}. That is what has to be abandoned.

\subsection{Inge Helland}\label{Inge}
\subsubsection{Introduction}
\label{intro}

In Section~\ref{Ch2}, it was argued that the three assumptions locality, realism and freedom together imply the CHSH inequality in Bell experiments. But loophole-free experiments have shown that this inequality may be violated in practice. Hence, one of the three assumptions needs to go. By the theory of relativity, I believe in the locality assumption. I also believe in the hypothesis of freedom/ independence, that is, independence between settings and parameters $\lambda_L , \lambda_X$ and $\lambda_Y$.

Thus, in this subsection, I will argue for a kind of non-realism, a non-realism connected to the variables that we take into account when we model future experiments. Specifically, I will show, using my own basic theory, that there can be no joint probability distribution for any output variables $X_1$, $X_2$, $Y_1$ and $Y_2$ in any model of a future run of a Bell experiment that is compatible with this theory.

From a general theorem on Hilbert space reconstruction, and from a natural interpretation of this theorem, it is proved that any actor might be limited in some specific sense during modelling. Explicitly, if he or she then considers two partially related maximal accessible theoretical variables during his/ her modelling of a future Bell experiment, he or she is not able to consider a third variable that is partially related to the first of these variables, but not to the second one.
It is argued from this that we have a situation where the CHSH inequality may be violated, given the assumptions of my theory.

The terms above will be precisely defined, and the consequences of this result will be thoroughly discussed below. In particular, the crucial terms maximal accessible variable and partially related pairs of variables will be explained.

The basic term in my theory will be (theoretical) variables $\theta , \lambda , \xi$ and $\eta$, where it is only assumed that if $\theta$ is a variable, and $\eta$ is a function of $\theta$, then $\eta$ is a variable. The variables may be accessible or inaccessible (to an actor or to a group of communicating actors), where again, it is assumed that if $\theta$ is accessible and $\eta$ is a function of $\theta$, then $\eta$ is accessible.

In the basic mathematical theory, the terms accessible and inaccessible are left undefined. But in the interpretation of the theory that leads to the quantum formalism, I need to assume that these terms are connected to a person or to a group of communicating persons. The single person is called Charlie below. Like Alice, Bob and Carol in the discussion of Section~\ref{Ch2}, Charlie can be assumed to be fictional.

But go back to the mathematical theory. A natural partial ordering among the variables is defined as follows: The variable $\eta$ is said to be less than or equal to the variable $ \theta$ if $\eta$ is a function of $\theta$, and it is assumed that maximal accessible variables exist with respect to this partial ordering. This means that if $\theta$ is maximal accessible, one cannot consider an accessible variable $\zeta$ and a non-injective function $f$ such that $\theta = f(\zeta)$. Informally, the maximal variable $\theta$ contains as much information as an accessible variable can contain.

Then, it is assumed that if $\lambda$ is any accessible variable, there exists a maximal accessible variable $\theta$ such that $\lambda\le \theta$, that is, $\lambda = f(\theta)$ for some function $f$. Two variables are defined to be equivalent if they are bijective/ one-to-one functions of each other. They then contain the same information.

In \citep{helland2026finalversionrecentapproach} it is proved that such simple assumptions lead to essential elements of the Hilbert space formalism of quantum mechanics if we look upon a situation with two non-equivalent maximal accessible variables taking a finite number of values. In \citet{helland2026bornrulenewquantum} it is shown that, if we in addition assume the validity of a version of the likelihood principle from statistics, Born's rule follows.

In the present subsection, I will not use this last result much, but merely take the other assumptions above as points of departure for discussing the Bell experiment. Note that in the qubit case, letting $\theta = X_a$ be the spin component in direction $a$, these assumptions can be seen as trivially satisfied. In the Bell experiment, all the variables $X_a$ and $Y_b$ may be seen as accessible variables. They are maximal accessible relative to a two-dimensional version of my main theorem. Note that in my theory, the notion `variables' does not necessarily mean random variables defined on one probability space.

In the Bell experiment, also a last assumption holds: There exists an inaccessible variable $\phi$ such that all accessible variables can be seen as functions of $\phi$. In the Bell experiment case, we can first let $\phi_1$ be a random radius vector in the Bloch sphere, and define $X_a = \mathrm{sign}(\mathrm{cos} (\phi_1 ,a))$. Note that the Bloch sphere here must be seen as a construction related to the mind of a person trying to model the experiment, and $\phi_1$ is inaccessible, it can be modelled as a random vector, but it takes no value outside the mind of the modeller.

Similarly, there exists an inaccessible variable $\phi_2$, the radius vector in a new Bloch sphere, such that we can define $Y_ b = \mathrm{sign}(\mathrm{cos} (\phi_2 ,b))$. Again, this can be seen as a random vector. Take $\phi = (\phi_1 , \phi_2)$.

A joint distribution of $\phi_1$ and $\phi_2$ would imply a joint probability distribution of $X_1 , X_2 , Y_1$ and $Y_2$, and by Theorem 4 below, no person is able to consider all these variables simultaneously during modelling, so one has difficulties in establishing such a joint probability distribution. From this, $\phi_1$ and $\phi_2$ seem not to have a joint distribution.

We can think of the frequency interpretation of probability. From a long series of repetitions in the Bell experiment, we can estimate one-dimensional and two-dimensional probabilities for the variables $X_1 , X_2 , Y_1$ and $Y_2$. But the sequence that could have been used to estimate a four-dimensional distribution, apparently does not converge.

For the Bell experiment, this model will be shown to imply that no actor is able to accept all the assumptions behind the CHSH inequality in a context where he or she is to make a model of the experiment.
The concrete mathematical result needed for this, is given by Lemma 1 in Subsection \ref{sec:10}, where these relations are explained in more detail.

All this relies on some purely mathematical results, where the Hilbert space formalism is derived from a few postulates.
The assumptions are related to the derivation of quantum theory, and as such can not be seen as `classical', as the assumptions of realism, locality and freedom in Section~\ref{Ch2}. However, in my opinion most of the assumptions made above, are quite intuitive, and it will be shown below that they may lead to violation of the CHSH inequality. Hence, I see these arguments as a way to find comfort with the results of the Bell experiments.

In Subsubsection \ref{subsec. 2}, the relevant actor/ person is called $C$, and his observation related to the EPR-Bohm experiment must be considered as explained by the same assumptions that lead to quantum mechanics. His observation is that spin components for Alice's particle and Bob's particle are opposite, even if Alice and Bob are far from each other.

In Subsubsection \ref{sec:10}, the relevant actor is called Charlie, and the conclusions that he arrives at during his attempts to model the Bell experiment, must be considered as real conclusions by other communicating people: During modelling, it is impossible to consider all the variables $X_1, X_2, Y_1$ and $Y_2$ at the same time defined on the same probability space. The strict assumption of realism must be rejected, and from this, the violation of the inequality can be understood.

I will start by discussing the EPR-Bohm experiment, which can be seen as a special case of a Bell experiment where Alice and Bob have decided to always use the same setting. The discussion here will be general, independent of my own Hilbert space theory in the form briefly described above and in more detail below. On the other hand, this Hilbert space theory will be crucial for my discussion of the ordinary Bell experiment.

This whole discussion can be seen as an update of the treatment in \citep{Helland2022a, Helland2023b}.

\subsubsection{Explaining the result of the EPR-Bohm experiment, assuming some quantum mechanics}
\label{subsec. 2}

In this Subsection, assume a joint entangled singlet state for the experiment
\begin{equation}
|\psi\rangle = (\sqrt{2})^{-1}(|1+, 2-\rangle - |1-, 2+\rangle ),
\label{s1}
\end{equation}
where spin components are assumed measured along some $z$-axis, and $|1+, 2-\rangle$ means that Alice's particle is in a state determined by that her spin component is $+1$, while Bob's particle is in the state where it is $-1$. Similarly, $|1-, 2+\rangle$ is determined.

As shown in \citep{Susskind_Friedman_2014}, this state is an eigenstate of the operator corresponding to the theoretical variable $\xi = \phi_1 \cdot \phi_2$ (see Theorem 1 below), where $\phi_1$ is the Bloch sphere spin vector for Alice's particle, $\phi_2$ denotes the Bloch sphere spin vector of Bob's particle, built from Pauli matrices, and $\cdot$ denotes the scalar product. The corresponding eigenvalue is $-3$.

This means the following: 1) Even though $\phi_1$ and $\phi_2$ are inaccessible variables, their dot product $\xi$ is in some way accessible to any observer $C$, given the joint state $|\psi\rangle$ for the experiment. 2) Specifically, $C$ is at any time forced to be in a state where, to him, $\xi = -3$. This must be true for any observer.

What does it mean that the scalar product is $-3$? Written in terms of the components along the $x$, $y$, and $z$-directions, we must have $\phi_1^x \phi_2^x + \phi_1^y \phi_2^y + \phi_1^z \phi_2^z = -3$. Since each of the spin components is $\pm 1$, each product here is $\pm 1$, and this is only possible if $\phi_1^x \phi_2^x = \phi_1^y \phi_2^y = \phi_1^z \phi_2^z = -1$. Thus, in the $x$-, $y$-, and $z$-directions, the spins are opposite, which implies that in any direction $a$, the spins must be opposite. Again, this is true for any observer, so, by a general epistemic interpretation of quantum mechanics, it can not be escaped from as a definite truth.

\subsubsection{A mathematical theory, and the theory related to a human actor}
\label{subsec. 3}

In general, let a person $C$ be in some context at the time t when he makes his observations or formulates his model of an experiment. In this context and at this time he can consider several variables: $\theta, \xi, \lambda, ...$. I will call these theoretical variables. If a variable $\lambda$ can be given some value at a future time, I will say that $\lambda$ is accessible. In the present treatment, it is enough to assume that all accessible variables take a finite number of values. 

Later, in connection to the Bell experiment, I will let the accessible variables be spin components $X_1 , X_2 , Y_1$ and $Y_2$ in a future run of the experiment, and functions of these variables.

The model below may on the one hand be seen as a purely mathematical model, where the terms theoretical variable and accessible variable are left undefined. On the other hand, it can as above be given a concrete interpretation related to a person or to some communicating persons. I will first formulate the mathematical model.

Say that a theoretical variable $\theta$ is `less than or equal to' the theoretical variable $\lambda$ if $\theta = f(\lambda)$ for some function $f$. This defines a partial ordering both among all theoretical variables and also among the accessible theoretical variables.

A basic assumption is that if $\lambda$ is accessible, and $\theta =f(\lambda)$ for some function $f$, then $\theta$ is also accessible.

One can also show that the spin components in different directions are \emph{related}. Then we first need the extra assumption that there exists an inaccessible variable $\phi$ such that all accessible variables can be seen as functions of $\phi$. In the spin component case, as in Subsection \ref{intro}, let $\phi$ be the spin vector in the Bloch sphere. By rotating this spin vector in the plane determined by the directions associated with two components $X_1$ and $X_2$, the two components can be made to coincide. 

In the general theory, if the two variables are $\theta$ and $\eta$, and a transformation in the $\phi$-space is denoted by $k$, then if for some function $f$, we have $\theta= f(\phi)$ and $\eta = f(k\phi)$, we say that $\theta$ and $\eta$ are related.

In the description above, and also below, every case where I introduce an observer $C$, this is merely meant to illustrate the theory, to have something concrete in our minds. But such a particular interpretation of the model will be important later. The theory itself is purely mathematical. About the accessible variables I assume as above that they are closed under functional dependence. An extra postulate is the existence of an inaccessible variable $\phi$ such that all the accessible ones are functions of $\phi$.

In general, an accessible variable is said to be maximal as such if it is maximal accessible with respect to the partial ordering above.
The existence of accessible variables that are maximal can either be motivated by Zorn's lemma, or stated as a separate postulate. 

In the general theory, I assume that if $\lambda$ is any accessible variable, then there exists a maximal accessible variable $\theta$ and a function $f$ such that $\lambda = f(\theta)$.

Assume now a future run of the Bell experiment. Let $X_a$ be the theoretical spin component of Alice's particle, measured in the direction $a$, and let $Y_b$ be the theoretical spin component of Bob's particle, measured in the direction $b$. As already stated, these may be looked upon as theoretical variables, and might as well be denoted by $\theta$ and $\eta$. They might or might not be related.

We are now in a special case of the following theorem, proved in \citep{Helland2024,helland2026finalversionrecentapproach} for $r\ge 3$, but also valid for $r=2$, cf. \citep{Helland2021,Helland2025JMP}. Later, in Subsubsection \ref{sec:10}, it will be used for $r=2$ and for $r=4$.
\bigskip

\textbf{Theorem 1.} \textit{In some context, let $\theta$ and $\eta$ be two non-equivalent maximal accessible theoretical variables, each taking $r$ values. Then, there exists an $r$-dimensional Hilbert space $\mathcal{H}$ such that each accessible variable $\zeta$ in the context has a self-adjoint operator $A^\zeta$ associated with it. In particular, there are self-adjoint operators $A^\theta$ and $A^\eta$.}

\textit{The operators $A^\zeta$ have in general the properties: 1) The eigenvalues of $A^\zeta$ are the possible values of $\zeta$. 2) The accessible variable $\zeta$ is maximal as such if and only if each eigenvalue of $A^\zeta$ is non-degenerate.} 

\bigskip

In the spin component case with $r=2$, we can take $\theta = X_1$ and $\eta = X_2$. Then $A^\theta$ and $A^\eta$ have non-degenerate eigenvalues $-1$ and $+1$, and the variables $X_1$ and $X_2$ are maximal as accessible variables. But note that we later also will discuss the case $r=4$, including also the outputs by Bob. From this perspective, $X_1$ and $X_2$ are no longer maximal. This is consistent with the fact that $A^\theta$ and $A^\eta$ now have degenerate eigenvalues. I will here consider maximal accessible variables of the form $C = (X_1 , Y_1)$, $D=(X_1 , Y_2)$, $E = (X_2 , Y_1)$, and $F = (X_2 , Y_2)$.

Connected to the property of relatedness, we also have the following result, again proved in \citep{Helland2024}:
\bigskip

\textbf{Theorem 2.} \textit{ If the maximal accessible $\theta$ and $\eta$ are related, there is a unitary operator $W=W(k)$, depending on the transformation $k$, such that}
\begin{equation}
A^\eta = W^\dagger A^\theta W.
\label{s2}
\end{equation}

Finally, there is an if and only if in Theorem 2; cf. \citep{Helland2023b,Helland2025JMP}:
\bigskip

\textbf{Theorem 3.} \textit{ Assume that for two maximal accessible variables $\theta$ and $\eta$, the relation (\ref{s2}) holds for an operator $W$. Then $\theta$ and $\eta$ are related.}
\bigskip

The following very important result follows immediately from these mathematical theorems:
\bigskip

\textbf{Corollary 1.} \textit{Assume that in some context, there are two different related maximal variables $\theta$ and $\eta$. Then it cannot in the same context and at the same time be a third maximal variable $\xi$ which is related to $\theta$, but not to $\eta$.}
\bigskip

Below, this Corollary will be connected to a concrete observer.
The qualification `at the same time' will then be important. By letting time vary, one can consider many variables, also unrelated ones. Also, different people may have very different notion of `maximal' in connection to their accessible variables; this may depend on their particular resources at the given time and in the given context.

\bigskip

\underline{Proof of Corollary 1.} 

First, we have by Theorem 2 that (\ref{s2}) holds. Assume that the maximal accessible $\theta$ and $\xi$ are related. Then, by the same Theorem, there is a unitary operator $V$ such that $A^\theta = V^\dagger A^\xi V$. By combining these two equations, we find $A^\eta = (VW)^\dagger A^\xi(VW)$. The operator $VW$ is unitary, so, it follows from Theorem 3 that $\eta$ and $\xi$ are related, which contradicts the assumptions made.
\qed
\smallskip

In the next Subsection, I want to specialise my theory to the Bell experiment. By the discussion in Section~\ref{Ch2}, under certain assumptions there are (stochastic) variables $X_a$ and $Y_b$ which denote the spin components of the two particles, with setting $a$ for Alice, and setting $b$ for Bob. In my discussion here, I will only consider the variables $X_a$ and $Y_b$ as accessible variables in the sense defined in the introduction \ref{intro}, but leave open the question whether all these variables are stochastic variables, defined on the same probability space. In agreement with my theory, the eventual measurements of $X_a$ or $Y_b$ can be done by any person $C$ (or by a communicating group of persons).

Now, include an additional general assumption of my theory: I assume that all of the theoretical variables, or at least some of them, the ones under consideration here, associated with some person $C$, can be seen as functions of an underlying inaccessible $\phi$, perhaps belonging to the subconsciousness of $C$. As such, $\phi$ can never be known by $C$, nor by any other person. In the purely mathematical version of the theory, I just assume that $\phi$ is inaccessible, and that all accessible variables can be seen as functions of $\phi$.

As a specific example, let $C$ consider the spin components of a particle. In this case, we can let $\phi$ be the inaccessible spin vector, an abstract vector such that the spin component in some direction $a$ is a discretised version of the projection of $\phi$ on $a$. In the binary case, this is just $X_a = \mathrm{sign}(\mathrm{cos}(a,\phi ))$. Concretely, in this case $\phi$ may be given by a vector in the well-known Bloch sphere.

An important remark must be made here: Past data are seen as real variables, but future data can be seen as theoretical variables that are represented in the mind of the actual person. These may be accessible or inaccessible to the person, depending upon what he can measure now or in the future.

It turns out, something which will be discussed in detail below, that a version of Corollary 1 will be important here. This corollary will limit which theoretical variables $C$ can consider during his or her modelling process, which models he or she is able to formulate and communicate to others. This, as it will turn out, will also be important on the question of what opinion $C$ will have about the CHSH inequality.

To some, it may be strange that a seemingly psychological result follows from pure mathematics. This may be explained by the notion of `accessible' for theoretical variables, which from a mathematical point of view can be seen as undefined in the Theorems above, but which, after the Theorems are proved, can be made precise in different directions, also a seemingly psychological one: It may be specified to mean that a given person $C$ is able to find a value for the variable in some given context. In connection to the Bell experiment, the context will be made precise below.

\subsubsection{The Bell experiment}
\label{sec:10}

The discussion in this Subsection will be in terms of a new person Charlie, who, like Alice, Bob, and Carol must be seen as a fictional person. The discussion could have been continued in purely mathematical terms, but the term `accessible variable' needs anyway reference to a person or to a group of communicating persons. The discussion will then also be easier to follow.

Assume that Alice and Bob meet after $n$ runs of the experiment and share their information on all the runs with the person Charlie. In concrete terms, Charlie has the following data: Settings for Alice in each run, $A=1\ \mathrm{or}\ 2$, Settings for Bob, $B=1\ \mathrm{or}\  2$, response $X$ for Alice ($X_1$ or $X_2$) and $Y$ for Bob ($Y_1$ or $Y_2$). We can imagine a situation where Charlie wants to make probability models related to a possible future new run of the experiment, and by the conditionality principle from statistics he will condition this analysis on the future $A$ and $B$.

 I do believe in the thesis that $(A, B$) is independent of what in Section~\ref{Ch2} is called $(\lambda_L , \lambda_X, \lambda_Y)$, so, in the notation of that section we have
 \begin{equation}
 \begin{array}{c}
 X_a = f(a, \lambda_L, \lambda_X)\\ Y_b = g (b, \lambda_L , \lambda_Y).
 \label{00}
 \end{array}
 \end{equation}
However, I will argue below that the variables $X_1, X_2, Y_1$ and $Y_2$ may not have a joint distribution, so then $\lambda_X$ and $\lambda_Y$ seem to be defined on different probability spaces. I will come back to that.
 
Charlie's analysis should at least be concentrated on the following parameters, related to the CHSH inequality, and estimable from the data sets that he has received:
\begin{equation}
l_1 =\mathbb{E}(XY|A=1,B=1 )=\mathbb{E}(X_1 Y_1),
\label{Bell6}
\end{equation}
\begin{equation}
l_2 =\mathbb{E}(XY|A=1,B=2 )=\mathbb{E}(X_1 Y_2),
\label{Bell7}
\end{equation}
\begin{equation}
l_3 =\mathbb{E}(XY|A=2,B=1 )=\mathbb{E}(X_2 Y_1),
\label{Bell8}
\end{equation}
\begin{equation}
l_4 =\mathbb{E}(XY|A=2,B=2 )=\mathbb{E}(X_2 Y_2).
\label{Bell9}
\end{equation}

Referring to Section~\ref{Ch2}, I make the assumption of freedom/ no-conspiracy here, so all the four expectations $\mathbb{E}(X_a Y_b)$ exist and satisfy the above equalities. The question is whether all these expectations can be related to the same probability space. The actual CHSH inequality is
\begin{equation}
-2 \le \mathbb{E} (X_1 Y_1) - \mathbb{E} (X_1 Y_2) - \mathbb{E} (X_2 Y_1) - \mathbb{E} (X_2 Y_2) \le 2.
\label{Bell2}
\end{equation}

Let us assume that for the purpose of his analysis of a future run, Charlie tries to make a probability model for the relevant variables. From the discussion of Subsubsection \ref{subsec. 3}, I will show that there is a limitation to how much Charlie is able to consider at some given time when making this model. His maximally accessible variables may be the different pairs $(\theta, \eta)$, where $\theta$ is a variable connected to Alice and $\eta$ is a variable connected to Bob.

Note that accessible variables either can be denoted by Latin letters or Greek letters. The discussion below will be in terms of Latin letters, the future spin components.

Consider then the four pairs $C=(X_1,Y_1)$, $D=(X_1, Y_2)$, $E=(X_2, Y_1)$ and $F=(X_2, Y_2)$. Every pair corresponds to one of the 4 parts of the data sets that are available to Charlie, and also a pair connected to the possible future run. 

In connection to such pairs, I will now define the new condition \emph{partially related}. I say that $(\theta , \eta )$ is partially related to $(\lambda , \xi )$ if either $\theta = \lambda$ and $\eta$ is related to $\xi$, or $\eta = \xi$ and $\theta$ is related to $\lambda$.

Specifically, $C$ and $D$ are partially related since they have a common $X_1$, and $Y_2$ may be made to coincide with $Y_1$ by a rotation. Similarly, $C$ and $E$ are partially related through a common $Y_1$. But $D$ and $E$ are not partially related.

Hence, while the earlier `related' notion was transitive, this new partial relation is not transitive.
\bigskip

\textbf{Lemma 1} \textit{ Corollary 1 has the following extension: Assume that in some context, there are two different partially related maximal variables $\alpha$ and $\beta$. Then it cannot in the same context and at the same time be a third maximal variable $\gamma$ which is partially related to $\alpha$, but not to $\beta$.}

\bigskip

\underline{On the proof of Lemma 1}

Consider the case where $\alpha = (X_1 , Y_1 )$ and $\beta =( X_1, Y_2 )$, and the case where $\alpha = (X_1 , Y_1 )$ and $\beta = (X_2 , Y_1)$. Apply Theorem 2 and Theorem 3 to the variables that can be rotated into each other. These are maximal accessible relative to a two-dimensional variant of Theorem 1. From these versions of Theorem 2 and Theorem 3, the modification of Corollary 1 follows.
$\qed$
\smallskip

I will also assume tentatively that the four-dimensional version of Theorem 1 can be extended to partially related variables. This will imply first that
according to this theorem, Charlie is then able to reconstruct a Hilbert space from the variables $C$ and $D$. Similarly, he will be able to construct a Hilbert space from $C$ and $E$. But the pairs $D$ and $E$ have no partial relationship to each other; they are in this sense essentially different. 

We are here in the situation of Lemma 1: Charlie is not able to consider all three pairs $C$, $D$ and $E$ when making decisions about the experiments. 

Specifically, the conditions of Lemma 1 hold for $\alpha=C$, $\beta=D$, $\gamma=E$.
Assume that Charlie considers these two variables. This implies that Charlie is not simultaneously able to consider the variable $E$, which is partially related to $C$, but not partially related to $D$. 

Charlie is thus in particular not able to put up a joint probability model for these 3 pairs. As a consequence of this, he is not able to put up a joint probability model for the 4 binary variables $X_1, X_2, Y_1$, and $Y_2$.

For assume, tentatively, that Charlie is able to put up a joint probability model for his four future variables $X_1, X_2, Y_1$, and $Y_2$. Then he would be able to deduce from this model also a joint probability model for the variables $C, D, E$ and $F$. Thus, from what has just been said, this tentative assumption is impossible. Charlie is therefore not able to think of a joint probability model for his four basic variables. In fact, he is not able to consider all these four variables when making decisions related to his probability model of the experiment.
\bigskip

\textbf{Theorem 4}

\textit{Assume my general theory. Then, in the Bell experiment, any person Charlie who tries to make a probability model of a future run, is not able to, at this time, consider simultaneously all the variables $X_1 , X_2 , Y_1$ and $Y_2$.}
\bigskip

Consider, as in Section~\ref{Ch2}, the simple argument behind the CHSH inequality. As discussed above, at this moment, he is not able to consider all the future theoretical variables $X_1, X_2, Y_1$ and $Y_2$ during his modelling. From this point of view, Charlie is simply not able to accept the assumptions leading to (\ref{Bell2}) when making his decisions. Thus, he cannot then accept the assumptions leading to the CHSH inequality, which contain consideration of future $X_1, X_2, Y_1$ and $Y_2$ during modelling. He must see the validity of this inequality either as an empirical question or as a question that can be resolved by his knowledge about quantum mechanics.

Concretely, consider the simple argument for the CHSH inequality based upon the fact that all the variables take the values $\pm 1$, so that
\begin{equation}
X_1Y_ 1 - X_1Y_2 - X_2Y_1 - X_2Y_2 =- X_2 (Y_1 + Y_2) + X_1 (Y_1 - Y_2) = \pm 2,
\label{Bell22}
\end{equation}
since one either has $Y_1 = Y_2$, so $Y_1 + Y_2 = \pm 2$ and $Y_1 - Y_2 =0$, or $Y_1 = -Y_2$, so $Y_1 - Y_2 = \pm 2$ and $Y_1 + Y_2 =0$.

Assuming that one can take expectation term for term in (\ref{Bell22}), CHSH of the form (\ref{Bell2}), follows. An equivalent assumption is that there is a single common hidden variable $\lambda$, defined on one probability space. Since the CHSH inequality may be violated, these assumptions must be wrong.

Concretely, these arguments violate the assumption of \emph{local realism}, the assumption that $(X_1, X_2, Y_1, Y_2 )$ can be commonly defined during his modelling process, that each $X_a$ and $Y_b$ take the values $\pm 1$; and that this vector has a distribution.

From his past data, Charlie can compute simple estimates: $\widehat{l_1}=\overline{X_1Y_1}$, $\widehat{l_2}=\overline{X_1Y_2}$, $\widehat{l_3}=\overline{X_2Y_1}$ and $\widehat{l_4}=\overline{X_2Y_2}$. Born's formula gives $\mathbb{E}(X_aY_b)=-\mathrm{cos}(a,b)$, where $a$ here is the direction of the setting chosen by Alice, and $b$ here is the direction of the setting chosen by Bob. Let us assume that the settings are such that the CHSH inequality is violated, say that the expression in the CHSH formula is $>2$. (One such choice is $a_1\sim 0^\circ$, $a_2\sim 90^\circ$, $b_1\sim 135^\circ$ and $b_2\sim 45^\circ$.) Then, if the number $n$ of runs is large enough, Charlie will find by using Born's formula before looking at his data, that with high probability from this:
\begin{equation}
\overline{X_1Y_1}-\overline{X_1Y_2}-\overline{X_2Y_1}-\overline{X_2Y_2} > 2.
\label{Bell10}
\end{equation}

Observing this will convince him that a version of the CHSH inequality is \emph{not} valid. He will then also predict that for a future series of runs, if the number of runs is large enough and the settings are as above, then (\ref{Bell10}) will hold with high probability.

Thus, he believes that the assumption of local realism must be violated. If he also is convinced about the validity of Einstein's relativity theory, he may deduce that the locality assumption should hold. Hence his only option is to reject the universal assumption of realism in the sense implied by Bell's theorem. Charlie may furthermore be convinced of the statement for instance advocated in \citep{Mermin1985} and \citep{Zwirn2020-ZWINVM-3}: The only `real things' in physics are events, and quantum theory should be connected to the perception of events made by an actor or by a group of communicating actors. Charlie as an actor is limited during the time where he works with his model, since he cannot put up a joint probability model for $X_1, X_2, Y_1$ and $Y_2$.

As discussed in Section~\ref{Ch2}, Bell in his later papers proved that local realism follows from an assumption of classical local causality. Hence, by the above reasoning, this assumption must also be rejected. 

\subsubsection{Discussion}
\label{sec:12}

Like Charlie, we are all limited. None of us can answer all questions, even simple questions that require a yes/ no answer. This is obvious, but our ability to answer questions may improve with time and experience, or by learning from others. However, it may not be so clear to everybody that parts of this limitation can be explained by the fact that we are not at the moment when we make our mental models, capable of keeping all relevant variables in our minds.

This particular limitation may be explained by referring to a version of Corollary 1, an immediate consequence of a mathematical theory based upon a few rather simple postulates, the same mathematical theory from which a foundation of the quantum formalism may be derived, and which also leads to a theory of decisions. 

These decisions can be made by a single person or by a group of communicating persons. Sometimes they are connected to measurements, and such measurements may be formalised simply by some given questions to nature, and sharp answers to such questions. The questions can be given in terms of what one can call accessible theoretical variables, which then give a connection to the quantum foundation proposed in \citep{Helland2023a,Helland2024,Helland2025JMP,helland2026finalversionrecentapproach,helland2026bornrulenewquantum}.

Partly, these arguments are based upon a general epistemic interpretation of quantum theory, but also we might say that the discussion around the actor Charlie and the eventual people that he communicates with has some universal flavour. From such arguments, and from the Bell theorem, we conclude that local realism cannot hold in general. In this subsection, that conclusion is based upon a concrete theory that implies our universal limitation in a modelling phase.

\subsection{Bart Jongejan}\label{Bart}
\subsubsection{Representation of Bell correlations with a complete weighted bipartite graph}\label{sec:graph}

I will present an alternative Bell-type thought experiment. Whereas Bell proposes that Alice and Bob each have two settings to choose between, I propose that each participant has many more settings to choose between. In fact, the more, the better (but also the longer a run of the experiment takes). The reason for this will become clear in this section. I will call this experiment `Carol's experiment'.
\begin{figure}[h]
	\centering
	\begin{subfigure}[h]{0.55\linewidth}
		\includegraphics[scale=0.5]{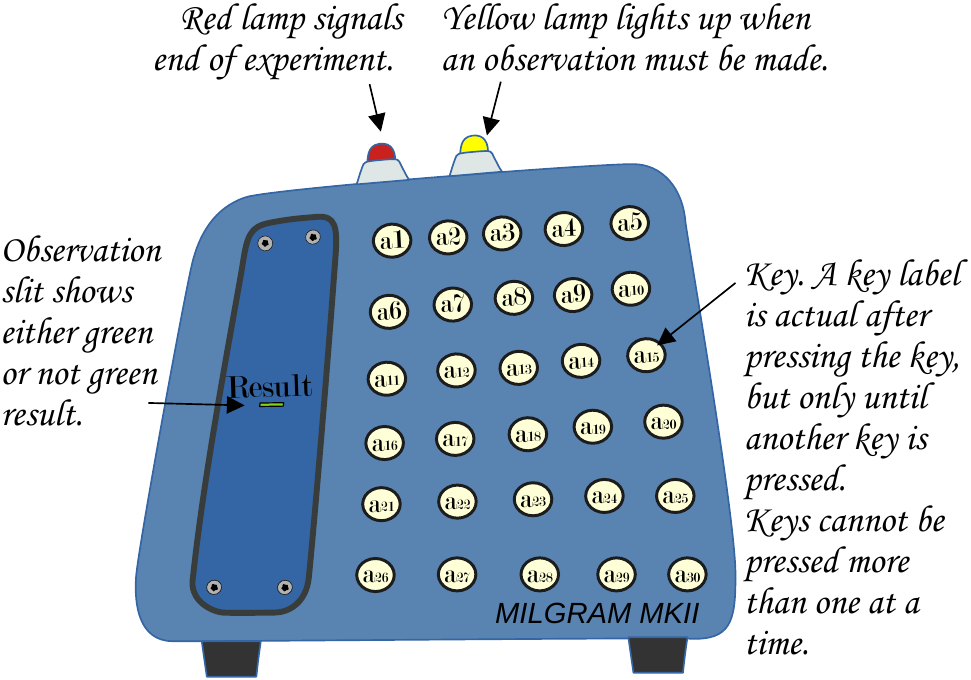}
		\caption{Alice's apparatus with indicator lamps, a narrow observation slit and keys $a_1$, $a_2$, \ldots, $a_{30}$}
		\label{fig:AliceInstrument}
	\end{subfigure}
	\hfill
	\begin{subfigure}[h]{0.34\linewidth}
		\raggedleft
		\includegraphics[scale=0.2]{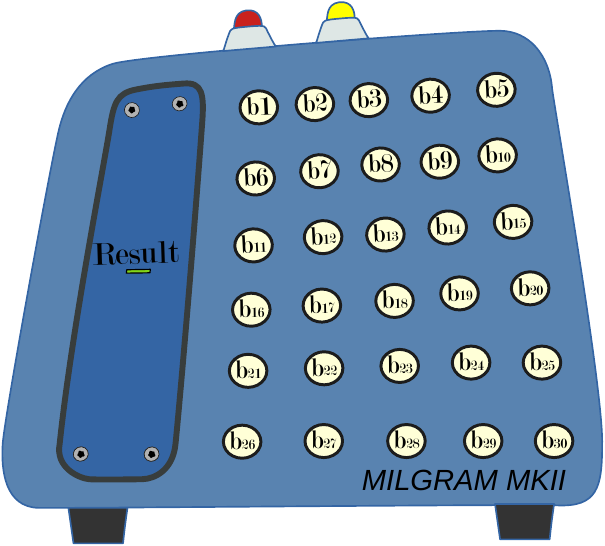}
		\caption{Bob's apparatus, similar to Alice's, with keys $b_1$, $b_2$, \ldots, $b_{30}$}
		\label{fig:BobsInstrument}
		\centering
		\includegraphics[scale=0.35]{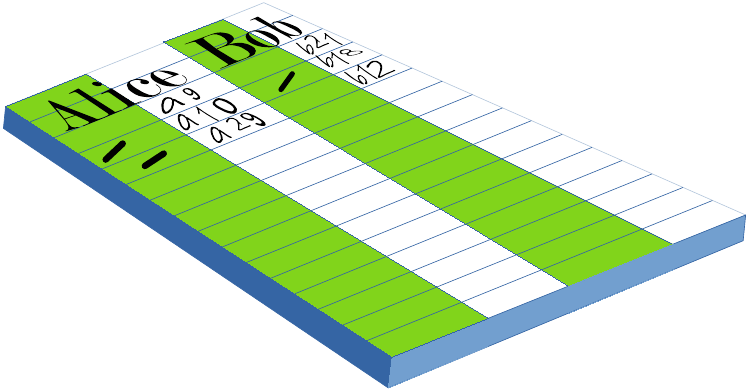}
		\caption{David's notepad with two columns for Alice's observations and key choices, and two columns for Bob's}
		\label{fig:Notepad}
	\end{subfigure}
	\caption{The essential hardware for Carol's experiment}
\end{figure}

Carol has enrolled two students in her experiment, Alice and Bob. Carol also has the help of David, who is a statistician. Alice and Bob are in separate rooms and cannot communicate with each other. There is also an apparatus in each room. See Figs. \ref{fig:AliceInstrument} and \ref{fig:BobsInstrument}. 
To mathematically describe the operational parts of the apparatuses, we need two sets of integers and two sets of labels. The two sets of integers are $\mathcal{K}=\left\{i | i \in \mathbb{N}, 1 \le i \le k, k \ge 2\right\}$ and $\mathcal{L}=\left\{j | j \in \mathbb{N}, 1 \le j \le l, l \ge 2\right\}$. The two sets of labels are $\mathcal{A}=\left\{a_i | i \in \mathcal{K}\right\}$ and $\mathcal{B}=\left\{b_j|j \in \mathcal{L} \right\}$.
Alice's apparatus has $k$ keys, uniquely labelled $a_1,a_2,\dots \in \mathcal{A}$. Bob's apparatus has $l$ keys, uniquely labelled $b_1,b_2,\dots\in \mathcal{B}$. To make the students believe they are participants in a psychological test, Carol tells them that the experiment is about reaction time. Their task is to press arbitrary keys, one at a time, and to quickly react when the yellow light on top of the apparatus turns on. Alice and Bob are encouraged to spread their key choices over all keys, but otherwise they are free to choose any key each time. Each time the yellow lamp lights up, which happens shortly after a key has been pressed, they must quickly check the narrow slit on the left-hand side of the front panel of their apparatus, before the light turns off again.

In the slit to the left of the keys, Alice and Bob sometimes observe a colour. We call that a $\textit{green}$ observation. At other times, the colour is absent; let us call that a $\textit{not-green}$ observation. In the long run, $\textit{green}$ and $\textit{not-green}$ observations occur equally often. Alice and Bob are not able to predict their outcomes; there is no regular pattern for them to discover.

After each observation, the student must send a telegraph message to David. The message must report which identifier was written on the pressed key and whether $\textit{green}$ or $\textit{not-green}$ was observed. 

David, who has a notepad with columns for Alice and Bob's observations (see Fig. \ref{fig:Notepad}), writes down the results in the order in which he obtains them\footnote{We assume that the experiment is performed in a loophole-free way. For example, Carol, independently of Alice, Bob and David, can check whether there is reason enough to qualify the students' observations as valid. If that is the case, she can send an `event ready' message to David, who will only accept observations when he has also received an `event ready' message from Carol.}. The students' measurements are events that are spacelike separated, so only a signal travelling faster than light would enable one of these events to exert an influence on the other event. After a very large number $N$ of trials ($N \gg |\mathcal{K} \times \mathcal{L}|$), the red lamps on Alice's and Bob's apparatuses turn on, signalling to them that they can stop pressing keys. David's notepad is now filled to the last line with results from $N$ trials.
Each of the four columns in the notepad can be regarded as a (discrete) variable. Alice's and Bob's key choices are two random variables $A\in \mathcal{A}$ and $B\in\mathcal{B}$, and their colour observations are two random variables $X\in\{-1,+1\}$ and $Y\in\{-1,+1\}$, where a \textit{green} observation is coded as $-1$ and a \textit{not{-}green} observation is coded as $+1$.

\begin{figure}
	\noindent\centerline{\includegraphics[width=0.8\textwidth]{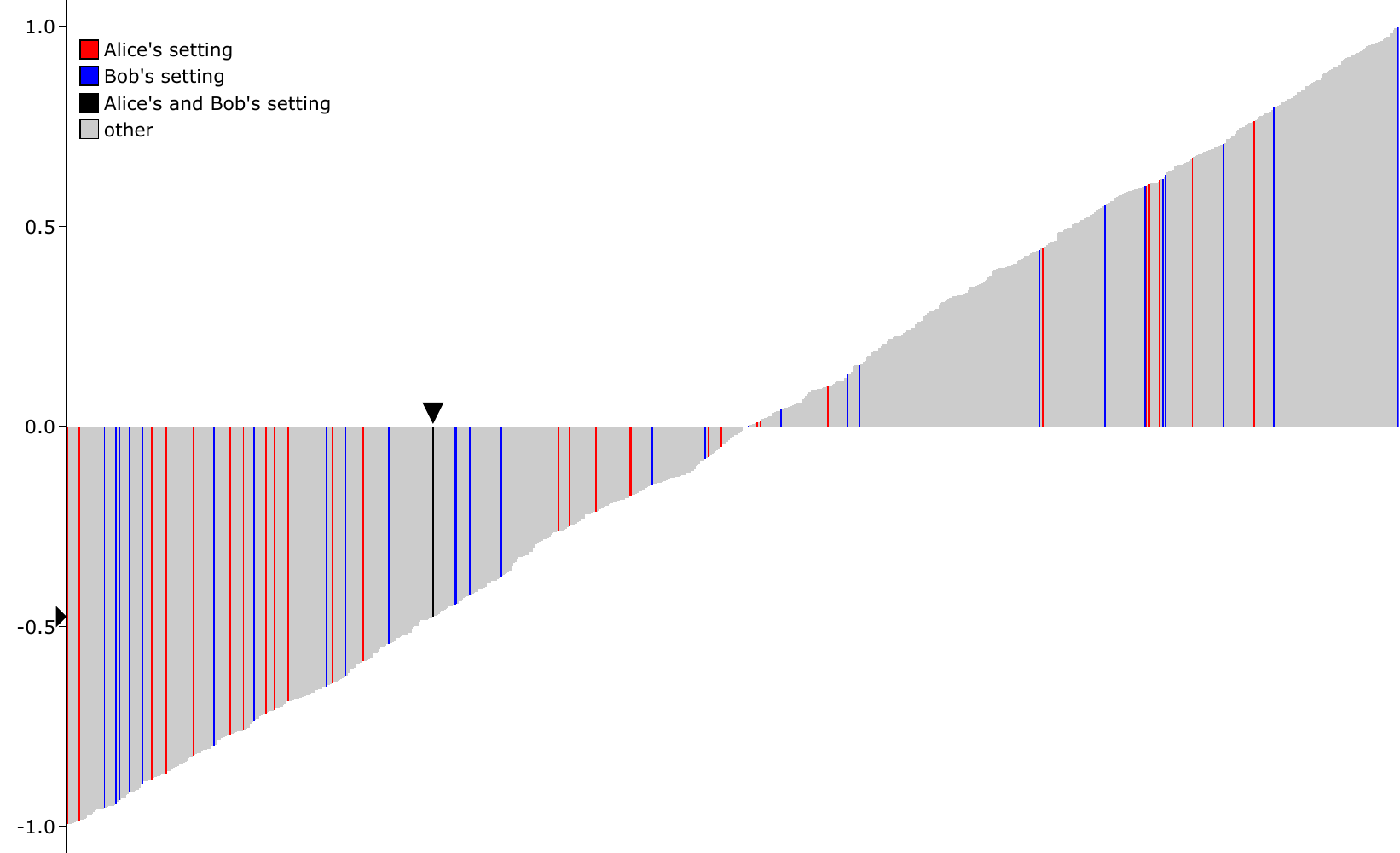}}
	\caption{Distribution of $\mathbb E_{ab}(X Y)$ for $ |\mathcal{K} \times \mathcal{L}| = 30\times 30$ settings. The distribution of $\mathbb E_{a_{10}b}(X Y)$ is in red. The distribution of $\mathbb E_{ab_{15}}(X Y)$ is in blue. The expectation value $\mathbb E_{a_{10}b_{15}}(X Y)$ is in black and its position and height are marked with black triangles. All other $841$ expectation values are in grey.}
	\label{fig:Sab-sorted-dim3-Alice30-Bob30}
\end{figure}

These are the results of the statistical analysis of David's data:
\begin{itemize}
	\item It is plausible that the random variables $\left(A,B,X,Y\right)$ are pairwise independent. That is, 
	\begin{equation}
		\begin{aligned}\label{prob1}
		 Prob\left(A=a\;\&\;B=b\right)=Prob\left(A=a\right)\times Prob\left(B=b\right)
		\end{aligned}
	\end{equation}
	and similarly for the pairs $\left(A,X\right)$, $\left(A,Y\right)$, $\left(B,X\right)$, $\left(B,Y\right)$, and $\left(X,Y\right)$.
	\item It is plausible that the random variables $\left(A,B,X\right)$ are mutually independent. The same is the case for $\left(A,B,Y\right)$, $\left(A,X,Y\right)$, and $\left(B,X,Y\right)$. Thus, 
	\begin{equation}
		\begin{aligned}\label{prob2}
			 Prob\left(A=a\;\&\;B=b\;\&\;X=x\right)=Prob\left(A=a\right)\times Prob\left(B=b\right)\times Prob\left(X=x\right)
		\end{aligned}
	\end{equation} etc.
	\item It is plausible that all four random variables are not mutually independent. That is,
	\begin{equation}
		\begin{aligned}\label{prob3}
			&Prob\left(A=a\;\&\;B=b\;\&\;X=x\;\&\;Y=y\right)\\
			\ne \ &Prob\left(A=a\right)\times Prob\left(B=b\right)\times Prob\left(X=x\right)\times Prob\left(Y=y\right)
		\end{aligned}
	\end{equation}
	The correlation between $X$ and $Y$ for a pair of keys $\left(a,b\right)$, can be quantified as follows:
	
	\begin{subequations}\label{correl}
		\begin{equation}
			\begin{aligned}\label{correlH}
				\mathbb E_{a b}(X Y) =
				&-P\left(X=-1\;\&\;Y=+1\mid A=a\;\&\;B=b\right)\\	
				&-P\left(X=+1\;\&\;Y=-1\mid A=a\;\&\;B=b\right)\\
				&+P\left(X=-1\;\&\;Y=-1\mid A=a\;\&\;B=b\right)\\
				&+P\left(X=+1\;\&\;Y=+1\mid A=a\;\&\;B=b\right)
			\end{aligned}
		\end{equation}
	\end{subequations}
	
	\item The set of all $\mathbb E_{ab}\left(XY\right)$ as $\left(a,b\right)$ varies over $\mathcal{A} \times \mathcal{B}$ is uniformly distributed over a finite set in the interval $\left[-1,1\right]$.
	\item For every $a \in \mathcal{A}$, the set of $\mathbb E_{ab}\left(XY\right)$ as $b$ varies over $\mathcal{B}$ is uniformly distributed over a finite set in the interval $\left[-1,1\right]$. See the red bars in Fig. \ref{fig:Sab-sorted-dim3-Alice30-Bob30}.
	\item For every $b \in \mathcal{B}$, the set of $\mathbb E_{ab}\left(XY\right)$ as $a$ varies over $\mathcal{A}$ is uniformly distributed over a finite set in the interval $\left[-1,1\right]$. See the blue bars in Fig. \ref{fig:Sab-sorted-dim3-Alice30-Bob30}.
	\item Define
	\begin{equation}
		\begin{aligned}\label{SBell}
			S_{pqrs}&=
			\mathbb E_{a_p b_q}(X Y)+
			\mathbb E_{a_r b_q}(X Y)+
			\mathbb E_{a_r b_s}(X Y)-
			\mathbb E_{a_p b_s}(X Y)
		\end{aligned}
	\end{equation}
	For all labels $a_p,a_r,b_q,b_s$
	\begin{equation}
		\begin{aligned}\label{SBell2}
			\abs{S_{pqrs}} &\le 2 \sqrt{2} \qquad \text{(Tsirelson's bound)}
		\end{aligned}
	\end{equation}
	\item For some choices of labels we have 
\begin{equation}\begin{aligned}\label{SBell3}
		\abs{S_{pqrs}}>2
	\end{aligned}
	\end{equation}
\end{itemize} 
There is a mathematical model that explains these results. The model functions at two levels. At the high level, the model treats the statistics of Carol's experiment as ground truths, while the low-level part explains how the statistics arise. First we discuss the high-level part.

Let $G=\left(\mathcal{A},\mathcal{B},\mathcal{C}\right)$ be a node labelled complete bipartite graph, where $\mathcal{C}=\left\{c_{a_ib_j}|i \in \mathcal{K}, j \in \mathcal{L} \right\}$ is the set of undirected labelled and weighted edges linking every element in $\mathcal{A}$ with every element in $\mathcal{B}$, such that the edge $c_{a_ib_j}$ links the node labelled $a_i$ with the node labelled $b_j$. 
Let the weights of the edges be real numbers $w\bigl(c_{a_ib_j}\bigr) \in \left[0,2\right]$.

Let
\begin{equation}
	\begin{aligned}\label{Sbip}
		S^{Graph}_{pqrs} &= 2- w\bigl(c_{a_pb_q}\bigr)-w\bigl(c_{a_rb_q}\bigr)-w\bigl(c_{a_rb_s}\bigr)+w\bigl(c_{a_pb_s}\bigr)\\
	\end{aligned}
\end{equation}
Then, by setting $w\bigl(c_{a_pb_q}\bigr)=w\bigl(c_{a_rb_q}\bigr)=w\bigl(c_{a_rb_s}\bigr)=0$
and $w\bigl(c_{a_pb_s}\bigr)=2$,
or by setting $w\bigl(c_{a_pb_q}\bigr)=w\bigl(c_{a_rb_q}\bigr)=w\bigl(c_{a_rb_s}\bigr)=2$ 
and $w\bigl(c_{a_pb_s}\bigr)=0$, it is easy to show that
\begin{equation}
	\begin{aligned}\label{Sbipsup}
		\sup \abs{S^{Graph}_{pqrs}} = 4
	\end{aligned}
\end{equation}
If the weights $w$ are random reals in the interval $\left[0,2\right]$ it is possible, though improbable, that there are four weights in a finite set of random weights to fulfil Eq. \eqref{Sbipsup}, 
but we will now see that there are random uniform distributions of real-valued weights $w$ in the interval $\left[0,2\right]$ that result in lower suprema of the absolute value of the $S$-expression \eqref{Sbip}.

Suppose that each element of $\mathcal{A}$ labels a random point on an abstract unit $n$-sphere\footnote{A 1-sphere is a circle, since a point on a circle can be uniquely represented by a single number, a 2-sphere is a sphere embedded in three-dimensional space, and so on.}. The labels $\mathcal{B}$ identify another set of random points on the same sphere. For Bell experiments with entangled spin $\frac{\hbar}{2}$ particles in the singlet state, using the notation $\angle\bigl(a,b\bigr)$ for the angle between the radial lines that connect the points $a$ and $b$ on the unit sphere to the centre of the sphere, we define a closeness angle $\eta\bigl(a_i,b_j\bigr)$ that is a linear function of the angle $\pi-\angle\bigl(a_i,b_j\bigr)$ between Alice and Bob's measurement directions. 
Define the weight $w\bigl(c_{a_ib_j}\bigr)$ as a function of $\eta$. Two conditions must be fulfilled by $w$. The first condition is that $w$ has a uniform distribution in the interval $\left[0,2\right]$, and the second condition is that $w\bigl(c_{a_ib_j}\bigr)$ increases monotonically with $\eta\bigl(a_i,b_j\bigr)$:
\begin{equation}
	\begin{aligned}\label{sepconda}
		\left(\forall p \in \mathcal{K}\right)
		\left(\forall q,s \in \mathcal{L}\right) 
		\eta\bigl(a_p,b_q\bigr) < \eta\bigl(a_p,b_s\bigr)
		\rightarrow
		w\bigl(c_{a_pb_q}\bigr) < w\bigl(c_{a_pb_s}\bigr)
	\end{aligned}
\end{equation}
and
\begin{equation}
	\begin{aligned}\label{sepcondb}
		\left(\forall p,r \in \mathcal{K}\right)
		\left(\forall q \in \mathcal{L}\right) 
		\eta\bigl(a_p,b_q\bigr) < \eta\bigl(a_r,b_q\bigr)
		\rightarrow
		w\bigl(c_{a_pb_q}\bigr) < w\bigl(c_{a_rb_q}\bigr)
	\end{aligned}
\end{equation}
The first condition is fulfilled if we require that the value of
$w\bigl(c_{a_pb_s}\bigr)$
is proportional to the number of edges $c_{a_pb_q}$ for which $w\bigl(c_{a_pb_q}\bigr)<w\bigl(c_{a_pb_s}\bigr)$. Now consider the cap of an $n$-sphere
$\leftidx{_n}D{}{}\left(a_p,b_s\right)$ that
has point $a_p$ at the apex and the point $b_s$ on the edge of the disk forming the base of the cap. This cap contains all points $b_q$ that are closer to the apex $a_p$ than $b_s$. Since the points are uniformly distributed, the expected number of points $b_q$ that are closer to the apex $a_p$ than $b_s$ is proportional to the area of the cap $\leftidx{_n}D{}{}\bigl(a_p,b_s\bigr)$. Thus, if we make $w\bigl(c_{a_pb_s}\bigr)$ proportional to the area of the cap, both the requirement that $w\bigl(c_{a_pb_s}\bigr)$ increases in proportion to the number of edges for which $w\bigl(c_{a_pb_q}\bigr)<w\bigl(c_{a_pb_s}\bigr)$ and the requirement that $w\bigl(c_{a_pb_s}\bigr)$ increases monotonically with $\eta\bigl(a_i,b_j\bigr)$ are fulfilled.
If we swap the roles of the points $a_p$ and $b_s$ and let $b_s$ be the apex of the cap, we find the same weight $w\bigl(c_{a_pb_s}\bigr)$, because the caps $\leftidx{_n}D{}{}\left(a_p,b_s\right)$ and $\leftidx{_n}D{}{}\left(b_s,a_p\right)$ have the same area. 

We will now construct a function of $\eta\bigl(a,b\bigr)$ that varies between 0 and 2 and does so in proportion to the area of $\leftidx{_n}D{}{}\left(a,b\right)$. Let $\tau$ be a shorthand for $\eta\bigl(a,b\bigr)$. 
For a positive constant number $C$, the area $\leftidx{_n}O{}{}\left(\tau\right)$ of the hyper-spherical cap $\leftidx{_n}D{}{}\left(a,b\right)$ is
\begin{equation}
	\begin{aligned}[b]\label{hypersphericalcap}
		\leftidx{_n}O{}{}\left(\tau\right) = C \int_{0}^{\tau} \sin[n-1](\theta) \ \dd{\theta} = C I\left(\tau,n\right)
	\end{aligned}
\end{equation}
\noindent Using integration by parts, we have, for $0 \le \tau \le \pi$ and $n\ge 1$:
\begin{equation}
	\begin{aligned}[b]\label{integral}
		I\left(\tau,n \right) =\int_{0}^{\tau} \sin[n-1](\theta) \dd{\theta}= \begin{cases}
			\tau & \text{if $n=1$}\\
			1 - \cos(\tau) \text{if $n= 2$}\\
			-\frac{1}{n-1} \sin[n-2](\tau) \cos(\tau) + \frac{n - 2}{n -1} I\left(\tau, n - 2\right) & \text{if $n\ge 3$}
		\end{cases}
	\end{aligned}
\end{equation}
\noindent Eq. \eqref{integral} computes a number that is proportional to the area of the $n$-hyper-spherical cap that contains all points that are closer to the apex point than $\tau$. If $M$ points are uniformly distributed over the $n$-sphere, then a fraction $\frac{I\left(\tau,n \right)}{I\left(\pi,n \right)} M$ is expected to lie within the $n$-hyper-spherical cap.

Define the function 
$\leftidx{_n}{H}{}\left(\upsilon\right)$, with $0 \le \upsilon\bmod{2\pi} \le \pi$,
as follows:
\begin{equation}
	\begin{aligned}\label{expectationValueA}
		\leftidx{_n}{H}{}\left(\upsilon\right) = 
		\frac{2 I\left(\upsilon\bmod{2\pi},n\right)}{I\left(\pi,n\right)}		 
	\end{aligned}
\end{equation}
$\leftidx{_n}{H}{}\left(\upsilon\right)$ is proportional to the number of random unit vectors that are closer to the apex than $\upsilon$. $\leftidx{_n}{H}{}\left(\upsilon\right)$ is uniformly distributed if $\upsilon$ is the closeness angle between unit vectors that are drawn at random.

\begin{figure}
	\noindent\centerline{\includegraphics[width=0.8\textwidth]{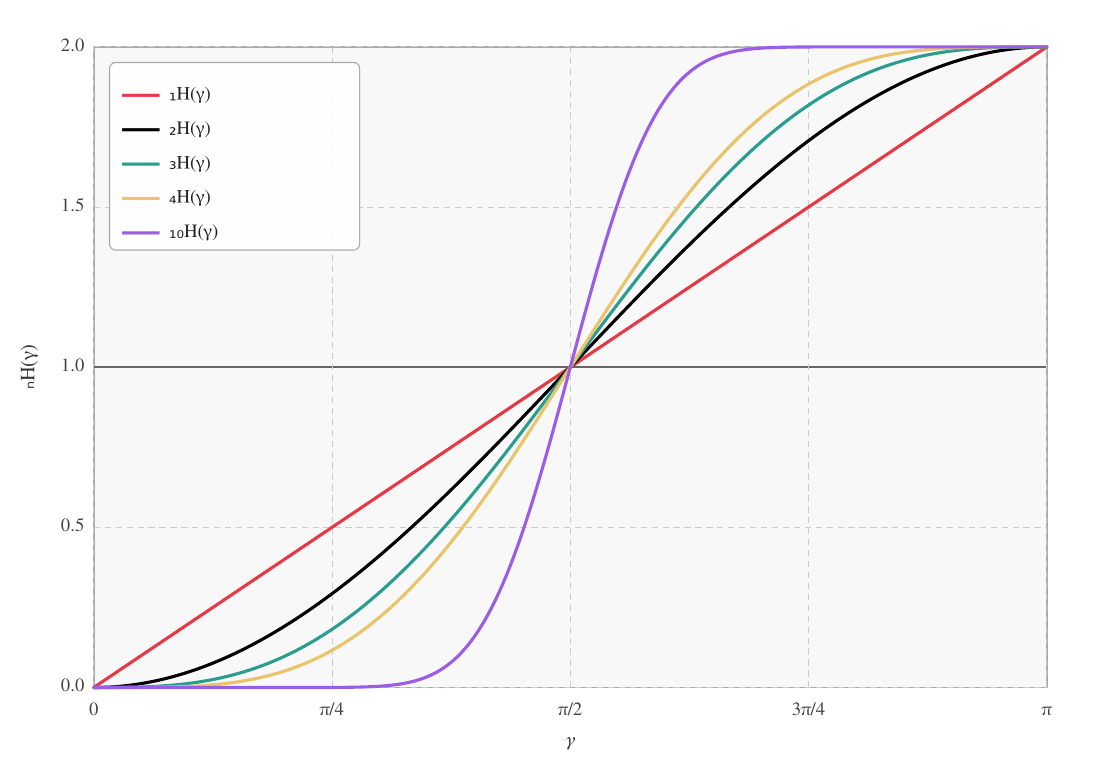}}
	\caption{$\leftidx{_n}{H}{}\left(\upsilon\right)$ for several values of $n$.}
	\label{fig:1H-20H}
\end{figure}

Eqs. \eqref{H1}, \eqref{H2}, \eqref{H3}, and \eqref{H4} are explicit expressions for $\leftidx{_n}{H}{}\left(\upsilon\right)$ for 2, 3, 4 and 5 spatial dimensions. See also Fig. \ref{fig:1H-20H}.
\begin{equation}
	\begin{aligned}\label{H1}
		\leftidx{_1}{H}{}\left(\upsilon\right) = \frac{2}{\pi}\left(\upsilon\bmod{2\pi}\right)
	\end{aligned}
\end{equation}
\begin{equation}
	\begin{aligned}\label{H2}
		\leftidx{_2}{H}
		{}\left(\upsilon\right) = 1-\cos(\upsilon)
	\end{aligned}
\end{equation}
\begin{equation}
	\begin{aligned}\label{H3}
		\leftidx{_3}{H}{}\left(\upsilon\right) = \frac{2}{\pi}\left(\left(\upsilon\bmod{2\pi}\right) -\sin(\upsilon)\cos(\upsilon)\right)
	\end{aligned}
\end{equation}
\begin{equation}
	\begin{aligned}\label{H4}
		\leftidx{_4}{H}{}\left(\upsilon\right) = 1-\cos(\upsilon)-\frac{1}{2}\sin[2](\upsilon)\cos(\upsilon)
	\end{aligned}
\end{equation}
Let us now associate four nodes labelled $a_p,b_q,a_r,b_s$ of the bipartite graph with four points on the unit sphere. Set
\begin{equation}
	\begin{aligned}\label{wasH}
		w\bigl(c_{a_pb_q}\bigr)=\leftidx{_n}{H}{}\left(\eta\bigl(a_p,b_q\bigr)\right)
	\end{aligned}
\end{equation} 
etc.
In order to compute the maximal violation of the CHSH expression\footnote{It is well known that the supremum of Eq. \eqref{Sbip} is reached when the first three $w$ terms on the right-hand side have the same value. It is also well known that the maximum violation of the CHSH inequality is not reached unless all four labels are represented by points on a great circle, thus making the closeness angle between the extreme points equal to three times the closeness angle between neighbouring points.}, with $0 \le \gamma \le \frac{\pi}{3}$ or $\frac{2\pi}{3} \le \gamma \le \pi$, set
\begin{equation}
	\begin{aligned}\label{gamma}
		\eta\bigl(a_p,b_q\bigr)=\eta\bigl(a_r,b_q\bigr)=\eta\bigl(a_r,b_s\bigr)=\gamma
		\qquad \text{and} \qquad 	
		\eta\bigl(a_p,b_s\bigr) = 3\gamma
	\end{aligned}
\end{equation} 
After substituting Eqs. \eqref{wasH} and \eqref{gamma} into Eq. \eqref{Sbip}, the CHSH expression, generalised to any number of spatial dimensions, becomes:
\begin{equation}
	\begin{aligned}\label{SbipH}
		\leftidx{_n}{S}{}\left(\gamma\right) &= 2-3 \times \leftidx{_n}{H}{}\left(\gamma\right)+\leftidx{_n}{H}{}\left(3\gamma\right)
	\end{aligned}
\end{equation}
Whereas Eq. \eqref{Sbip} leads to a supremum of $4$, the supremum of the absolute value of $\leftidx{_n}{S}{}\left(\gamma\right)$ is lower than that, and depends on the dimensionality of the unit sphere. This is seen as follows.
\noindent $|\leftidx{_n}{S}{}\left(\gamma\right)|$ reaches stationary points at $\gamma = \frac{\pi}{4}$ and $\gamma = \frac{3\pi}{4}$ for all values $n > 1$ because
\begin{equation}
	\begin{aligned}\label{chshma}
		&\diff{}{\gamma}\left[2-3\times \leftidx{_n}{H}{}\left(\gamma\right) + \leftidx{_n}{H}{}\left(3 \gamma\right)\right]
		\propto \diff{}{\gamma}\left[2-3\times I\left(\gamma,n \right) + I\left(3 \gamma\bmod{2\pi},n \right)-I\left(\pi,n\right)\right]\\
		\propto &\ -3 \sin[n-1](\gamma) + \diff{3\gamma}{\gamma}\sin[n-1](3\gamma)
		\propto -\sin[n-1](\gamma) + \sin[n-1](3\gamma)
	\end{aligned}
\end{equation}
and
\begin{equation}
	\begin{aligned}\label{chshm}
		&-\sin[n-1](\frac{\pi}{4}) + \sin[n-1](\frac{3\pi}{4})
		=-\sin[n-1](\frac{3\pi}{4}) + \sin[n-1](\frac{9\pi}{4})\\
		=&-\left(\sqrt{\frac{1}{2}}\right)^{n-1} + \left(\sqrt{\frac{1}{2}}\right)^{n-1}
		=\qq 0
	\end{aligned}
\end{equation}
The endpoints give smaller values.
Table \ref{tab:chshmx} shows the value of $\left|\leftidx{_n}{S}{}\left(\frac{\pi}{4}\right)\right|$ for some values of $n$.
\begin{table}
	\centering
	\begin{tabular}{ c c c c c c}
		$n$ & $\left|\leftidx{_n}{S}{}\right|_{\max}$ & $n$ & $\left|\leftidx{_n}{S}{}\right|_{\max}$ & $n$ & $\left|\leftidx{_n}{S}{}\right|_{\max}$\\
		1 & 2.000000 & 5 & 3.697653 &  9 & 3.940175 \\
		2 & 2.828427 & 6 & 3.800699 & 10 & 3.959522 \\
		3 & 3.273240 & 7 & 3.867418 & 11 & 3.972511 \\
		4 & 3.535534 & 8 & 3.911184 & 20 & 3.999066 \\
	\end{tabular}
	\caption{The maximum value of $\left|\leftidx{_n}{S}{}\right|_{\max}$ for some values of $n$}
	\label{tab:chshmx}
\end{table}

By setting the weights on the edges $c_{a_pb_q}$ equal to $\leftidx{_n}{H}{}\left(\eta\bigl(a_p,b_q\bigr)\right)$
we have created a complete bipartite graph with random weights on the edges $c_{a_pb_q}$ that have a uniform distribution in $\left[0,2\right]$, whether varying over both $p$ and $q$ or over either one while keeping the other fixed. Of particular importance is the value of $\leftidx{_2}{S}{}\left(\frac{\pi}{4}\right)$, which is Tsirelson's bound $2\sqrt{2}$. If $n=2$, the model violates the CHSH inequality by the same amount as QM.

If we associate each of Alice's key labels in Carol's experiment with a node $\in \mathcal{A}$, each of Bob's key labels with a node $\in \mathcal{B}$, and each expectation value $\mathbb E_{a_pb_q}\left(XY\right)$ with an edge $c_{a_pb_q}$ such that the expectation value is $+1$ when the edge has zero weight, then
\begin{equation}
	\begin{aligned}\label{Eisw}
		\mathbb E_{a_pb_q}\left(XY\right) =1-w\bigl(c_{a_pb_q}\bigr)= 1-\leftidx{_2}{H}{}\left(\eta\bigl(a_p,b_q\bigr)\right)
	\end{aligned}
\end{equation}
Whereas the abstract geometric approach does not point to a particular physical process by which the weights $w$ arise, Carol's experiment would be a physical process. The rightmost member of Eq. \eqref{Eisw} is an abstract mathematical expression, whereas the leftmost member, in principle at least, can be established experimentally by averaging over a series of observations. However, neither member of Eq. \eqref{Eisw} explains what happens in every single trial. That is the task of the lower level part of the proposed model.

In this section we have seen that the statistical data produced by a run of Carol's experiment can be modelled as geometric relations between two sets of random points on an abstract unit sphere. We have generalised the model to any number of spatial dimensions, and found that the case of a 2-dimensional sphere in 3d space fits the data from a run of the experiment. The weights are not fitted one by one: all of them must be realised by a single configuration of points on one sphere, whose number of dimensions is the only global parameter.

\subsubsection{One-dimensional hidden variables}

\begin{figure}[h]
	\centering
	\includegraphics[scale=0.5]{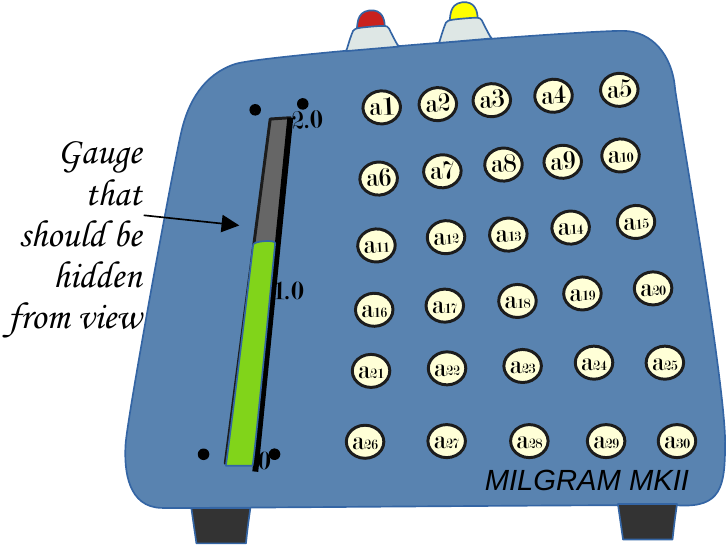}
	\caption{Alice's apparatus, cover over the fluid level gauge removed}	\label{fig:AliceInstrumentShowingGage}
\end{figure}
Let us return to Alice's and Bob's apparatuses and remove the covers over the slits that sometimes displayed a green colour and sometimes didn't. Fig. \ref{fig:AliceInstrumentShowingGage} shows Alice's apparatus without the cover. Behind the cover is a glass pipe filled with a green fluid. There is a scale along the full length of the glass pipe, running from $0.0$ at the lowest level via $1.0$ in the middle to $2.0$ at the top of the glass pipe. If Alice's fluid level $\lambda_A$ or Bob's fluid level $\lambda_B$ is above the $1.0$ mark, the colour visible through the slit in the cover would have been green. Otherwise, one would just see the indistinct background behind the glass. The pair $\lambda_A,\lambda_B$ is the hidden variable in this experiment.

The slits of Alice's and Bob's apparatuses implement step functions $f$ and $g$ of the green fluid levels $\lambda_A$ and $\lambda_B$ as in Eq. \eqref{gAG}.
\begin{equation}
	\begin{aligned}\label{gAG}
		X=f\left(\lambda_A\right)=
		\begin{cases}
			+1 &\text{if $0 \le \lambda_A \le 1$ (\textit{not{-}green}})\\
			-1 &\text{if $1 < \lambda_A \le 2$ (\textit{green})} \\
		\end{cases}\\
		Y=g\left(\lambda_B\right)=
		\begin{cases}
			+1 &\text{if $0 \le \lambda_B \le 1$ (\textit{not{-}green})}\\
			-1 &\text{if $1 < \lambda_B \le 2$ (\textit{green})}\\
		\end{cases}\\
	\end{aligned}
\end{equation}
Notice the similarities and differences between Eq. \eqref{gAG} and the Bell expression Eq. \eqref{BellEq13}.
\begin{equation}
	\begin{aligned}\label{BellEq13}
		X\left(\hat{a},\lambda\right)=-Y\left(\hat{a},\lambda\right)
	\end{aligned}
\end{equation}
The function $f$ in Eq. \eqref{gAG} has only one parameter and is in coordinate-free form. Bell's expression has two parameters, $\hat{a}$ and $\lambda$; if $\lambda$ is modelled as a classical spin vector, it invites the use of coordinates:
\begin{equation}
	\begin{aligned}\label{coord}
		f'\bigl(\hat{a},\hat{\lambda}\bigr)&=\sign\bigl(\hat{a}.\hat{\lambda}\bigr)=\sign\left(a_x\lambda_x+a_y\lambda_y+a_z\lambda_z\right)\\				g'\bigl(\hat{b},\hat{\lambda}\bigr)=-f'\bigl(\hat{b},\hat{\lambda}\bigr)&=\sign\bigl(\hat{b}.\bigl(-\hat{\lambda}\bigr)\bigr)=\sign\bigl(-b_x\lambda_x-b_y\lambda_y-b_z\lambda_z\bigr)
	\end{aligned}
\end{equation}

Let us ask Alice and Bob to perform another $N \gg |\mathcal{K}\times\mathcal{L}|$ observations. 
Instead of placing check marks in some fields in the green columns of his notepad, David now writes numbers between $0.0$ and $2.0$ in all green column fields as indications of the level of the green fluid in Alice's and Bob's apparatuses.

\begin{figure}[ht]
	\centering
	\includegraphics[scale=0.5]{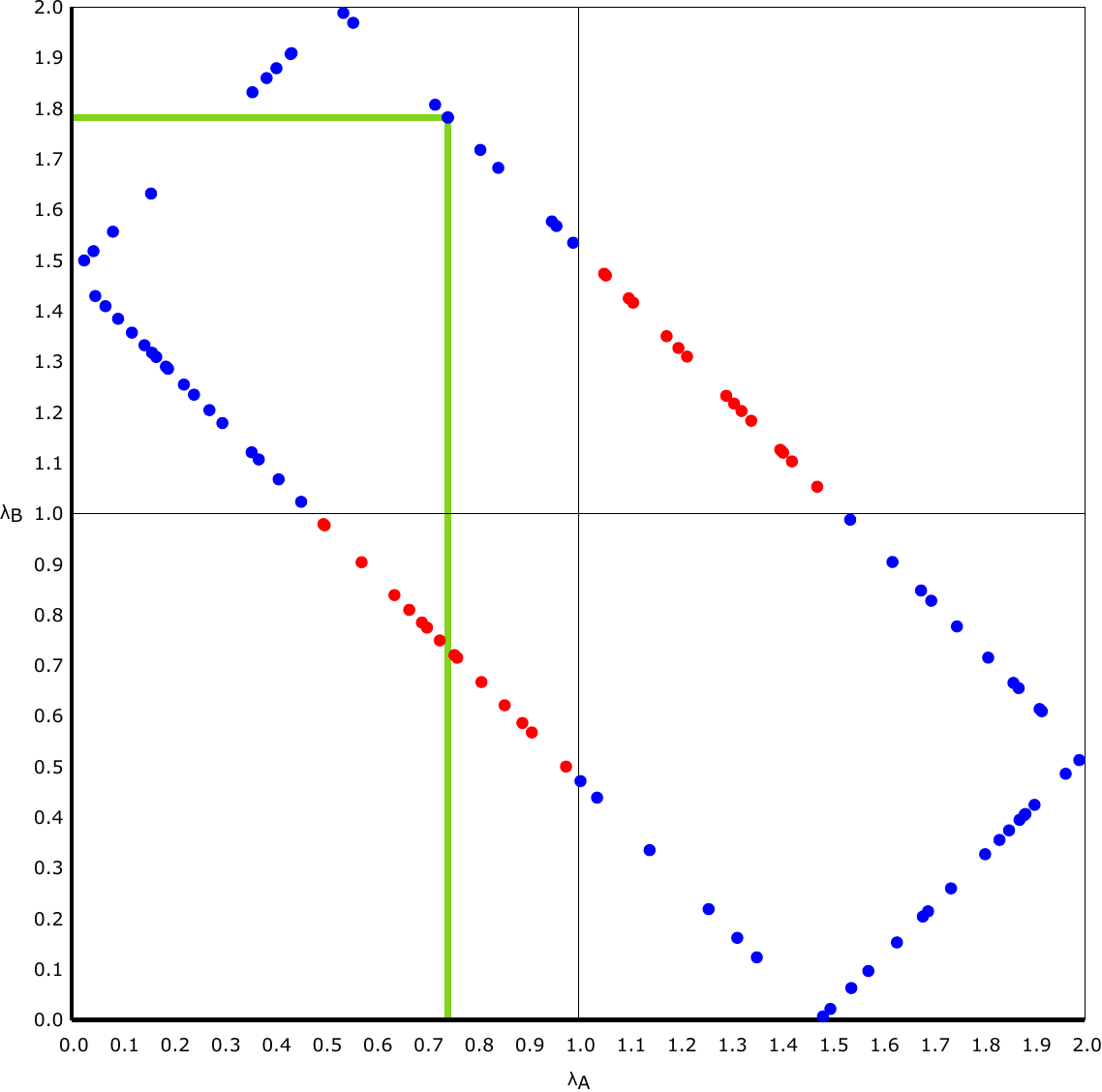}
	\caption{The values of $\lambda_A$ and $\lambda_B$ for the first hundred trials in an imaginary run of Carol's experiment, for key choices $\left(a_{10},b_{15}\right)$. The
		horizontal and vertical green lines represent Alice's and Bob's green fluid columns observed during a single trial. Red dots (lower left and upper right quadrant) are trials where Alice and Bob would have made the same colour observations if the covers still had been in place, blue dots (upper left and lower right quadrants) are trials where Alice and Bob would have made different binary observations. \label{fig:redblue}}
\end{figure}

David notices that all four variables in combination, the two fluid levels and the two key choices, are strongly correlated. Fig. \ref{fig:redblue} plots the first hundred values of $\lambda_A$ and $\lambda_B$ in a simulated test for a particular, but arbitrary, pair of keys $\left(a_{10},b_{15}\right)$.

Each dot in Fig. \ref{fig:redblue} represents the levels of two columns of green fluid. For example, there is a dot where Alice's green fluid level was $\lambda_A=0.742349$ and Bob's $\lambda_B=1.781968$. The horizontal green line depicts Alice's green fluid $\lambda_A$. The longer vertical green line depicts Bob's green fluid level $\lambda_B$. 

The dots are, without exception, restricted to the perimeter of a rectangle. It is almost as though Alice, if she knew her own green fluid level and both her own and Bob's key choices, would be able to predict or even determine Bob's green fluid level, but this is never the case, except when her green fluid level is exactly $0$ or $2$. In those two cases, Bob's green fluid level is predictable with certainty. In all other cases, there are two possibilities, and the probability that Alice guesses the correct height of Bob's green fluid is only $50\%$.

For key pairs other than $\left(a_{10},b_{15}\right)$, the diagram looks similar, but with different proportions between adjacent sides. The corners of each rectangle always touch the sides of a square spanned by the points $\left(0,0\right)$, $\left(2,0\right)$, $\left(2,2\right)$ and $\left(0,2\right)$, but the points where the rectangle touches the surrounding depends on the actual pair of key choices.

The red dots in Fig. \ref{fig:redblue} denote trials where Alice and Bob see the same colour in the slits of their apparatuses. So, Alice and Bob either both see no colour (lower left quadrant) or they both see a green colour through the slits of their apparatuses (upper right quadrant). A blue dot indicates that Alice and Bob made different observations. In the upper left quadrant are those observations where Bob observes $\textit{green}$ and Alice observes $\textit{not-green}$. The observations in the lower right quadrant are those where Alice observes $\textit{green}$ and Bob observes $\textit{not-green}$.

To compute the expectation value $\mathbb E_{a_{10} b_{15}}(X Y)$, we must count the number of trials in the red fields, subtract the number of trials in the blue fields, and divide the result by the number of trials with this key pair. See Eq. \eqref{expvalueObs}.
\begin{equation}
	\begin{aligned}\label{expvalueObs}
		\mathbb E_{a_i b_j}(X Y) =&\frac{\sum_{t=1}^N f\left(\lambda_A \left(t\right)\right) g\left(\lambda_B \left(t\right)\right)\delta_{a\left(t\right)a_i}\delta_{b\left(t\right)b_j}}{\sum_{t=1}^N \delta_{a\left(t\right)a_i}\delta_{b\left(t\right)b_j}}
	\end{aligned}
\end{equation}
The trials are randomly distributed along the perimeter of the rectangle, with a uniform density. As the number of trials increases, the result is approximated by integration of $f\left(\lambda_A\right)g\left(\lambda_B\right)$ along the perimeter of the rectangle over all allowed combinations of levels of green fluid $\left(\lambda_A,\lambda_B\right)$ while keeping the key choices $\left(a_i,b_j\right)$ fixed. By doing this integration for all combinations of keys, we find all values of the expectation value $\mathbb E_{a_i b_j}(X Y)$. 

\begin{figure}[h]
	\centering
	\begin{subfigure}[h]{0.6\linewidth}
		\centering
		\includegraphics[scale=0.3]{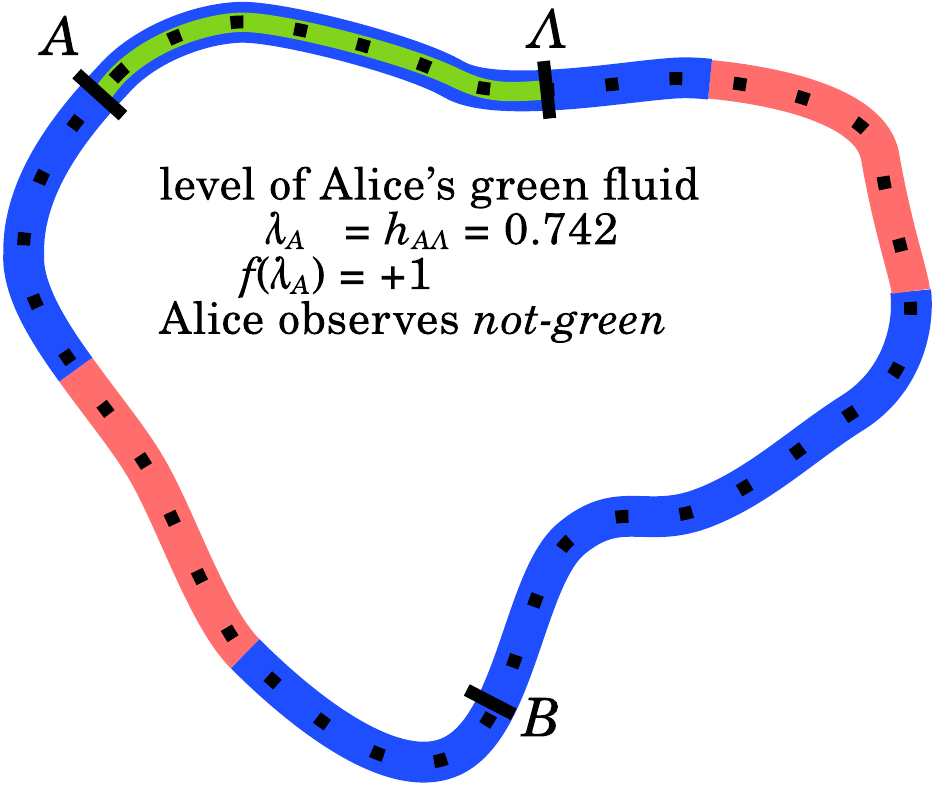}
		\caption{Alice's green fluid level\label{fig:contourgreenAliceBoxed}}
		\includegraphics[scale=0.3]{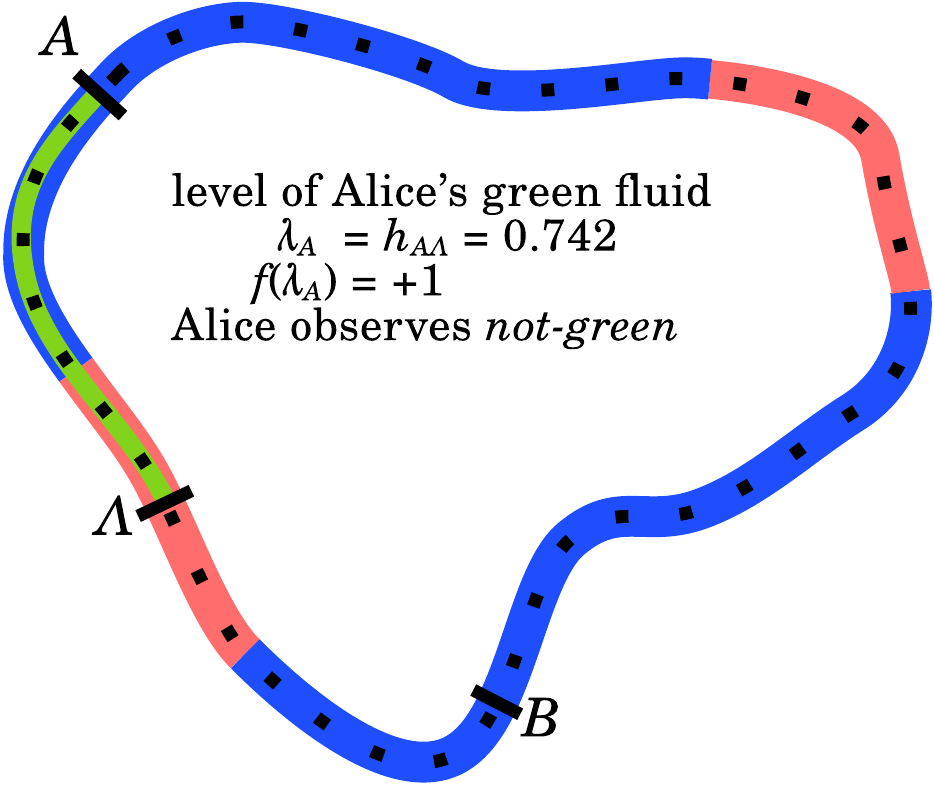}
		\caption{Alice, same green fluid level as in Fig. \ref{fig:contourgreenAliceBoxed}\label{fig:contourgreenAlicef2-boxed}}
	\end{subfigure}
	\hfill
	\begin{subfigure}[h]{0.35\linewidth}
		\centering
		\includegraphics[scale=0.3]{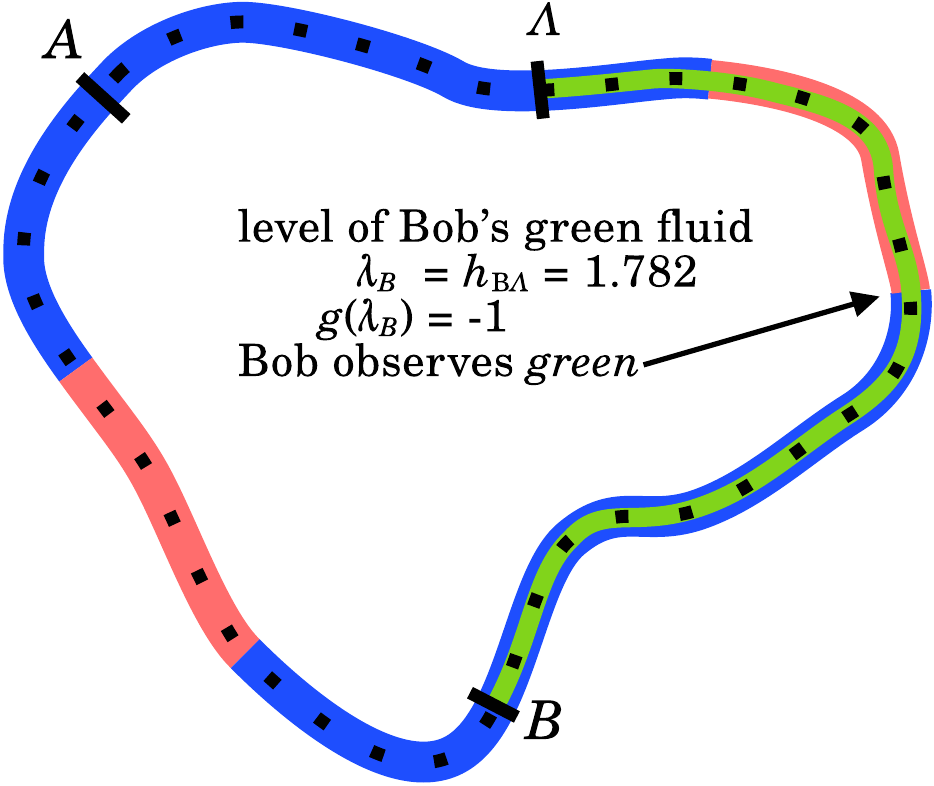}
		\caption{Bob's green fluid level $>1$\label{fig:contourgreenBobBoxed}}
		\includegraphics[scale=0.3]{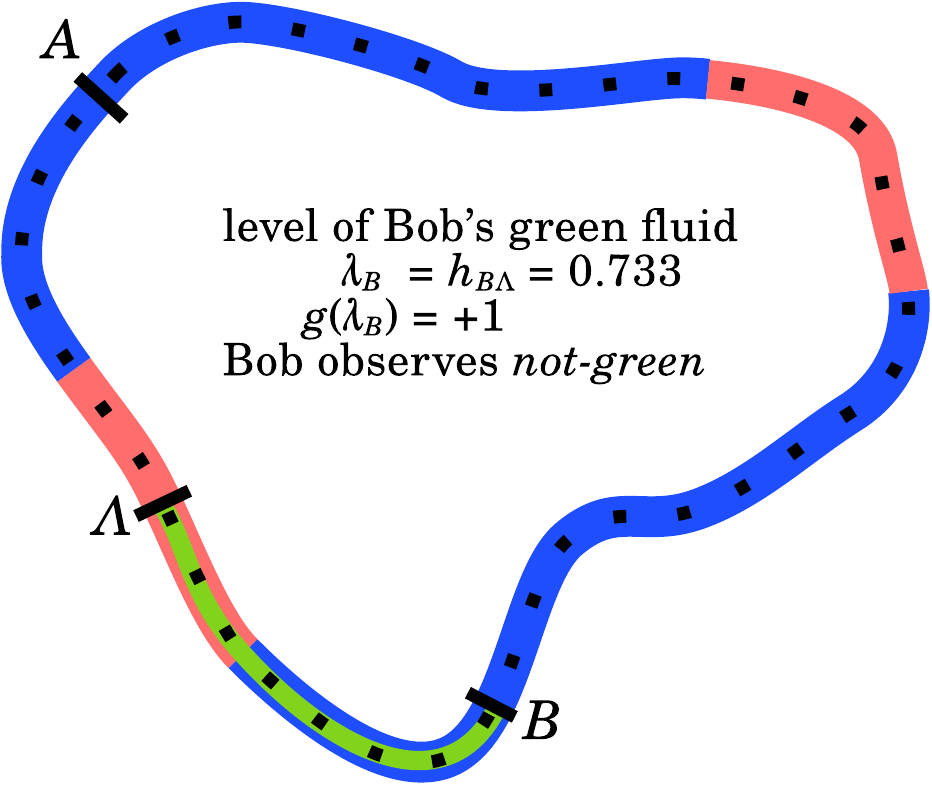}
		\caption{Bob's green fluid level $<1$\label{fig:contourgreenBobBoxed2}}
	\end{subfigure}	\caption{Observations by Alice and Bob as explained by hidden parameters $\lambda_A$ and $\lambda_B$.
	Alice and Bob make the same observation if $\Lambda$ is in a red sector, as in (\ref{fig:contourgreenAlicef2-boxed},\ref{fig:contourgreenBobBoxed2}), and different observations if $\Lambda$ is in a blue sector, as in (\ref{fig:contourgreenAliceBoxed},\ref{fig:contourgreenBobBoxed}). 
	These two pairs illustrate that even if Alice knows her own green fluid level and Bob's setting, she cannot predict with certainty the green fluid level of Bob.
	The small black squares are a linear graduation.
	The separation between two squares is $1/40^{\text{th}}$ of the perimeter.
		\label{fig:contourgreen}}
\end{figure}

Fig. \ref{fig:contourgreen} depicts the same loop four times. The loop has a perimeter of 4 and therefore we call it a `4-loop'.
The 4-loop does not add anything new with respect to the rectangle in Fig. \ref{fig:redblue}, but is different in one important detail. I will come back to that later. There are three points $\Lambda$, $A$, and $B$ on the 4-loop. The points $\Lambda$ and $A$ define the hidden variable $\lambda_A$ and the points $\Lambda$ and $B$ define the hidden variable $\lambda_B$. The point $\Lambda$ itself is not a hidden variable.
Points against a red background in Fig. \ref{fig:contourgreen} are either farther from both $A$ and $B$ than 1 or closer to both $A$ and $B$ than 1. For the points with a blue background, the opposite is the case.
In each instance of a 4-loop, the length of the green segment is equal to the height of the column of green fluid in the corresponding apparatus: $A$ symbolises Alice's setting and $B$ Bob's.

\noindent Fig. \ref{fig:contourgreen} resembles Fig. \ref{fig:redblue} in several ways.
\begin{itemize}
	\item 
	The rectangle in Fig. \ref{fig:redblue} corresponds to the 4-loop in Fig. \ref{fig:contourgreen}.
	\item The blue dot where the two green line segments meet in Fig. \ref{fig:redblue} corresponds to the point $\Lambda$ in Fig. \ref{fig:contourgreen}. The left and bottom corners of the rectangle in Fig. \ref{fig:redblue}, where Alice's and Bob's green fluid levels are zero, respectively, correspond to the points $A$ and $B$ in Fig. \ref{fig:contourgreen}.
	\item Red and blue dots in Fig. \ref{fig:redblue} correspond to points against a red and blue background, respectively, in Fig. \ref{fig:contourgreen}.
	\item 
	Horizontal and vertical green lines no longer than $2$ can be drawn from the vertical and horizontal axes to every point of the rectangle in Fig. \ref{fig:redblue}. Any two points in Fig. \ref{fig:contourgreen} can be connected by a curve no longer than $2$ that follows the outline of the 4-loop.
\end{itemize}
But there are also differences.
\begin{itemize}
	\item 
	The rectangle in Fig. \ref{fig:redblue} has a perimeter of $4\sqrt{2}$. The loop in Fig. \ref{fig:contourgreen} has a perimeter of $4$. 
	\item 
	The green lines in Fig. \ref{fig:redblue} cross the 2-dimensional plane, while the green curves in Fig. \ref{fig:contourgreen} stay within the confines of the 4-loop, which is 1-dimensional. 
	\item 
	In Fig. \ref{fig:redblue}, each dot is specified by two numbers, the green fluid levels of Alice and Bob. In Fig. \ref{fig:contourgreen}, every point has only a single coordinate, since a 4-loop is 1-dimensional. This is the important detail mentioned above.
\end{itemize} 
We can compute the expectation value $\mathbb E_{a_i b_j}(X Y)$ by associating settings $a_i$ and $b_j$ with the points $A$ and $B$, computing colour observations $X$ and $Y$ from the position of $\Lambda$ relative to $A$ and $B$, and taking the average of $XY$ while moving $\Lambda$ around the 4-loop. 
For that, we define a coordinate system on the 4-loop from an arbitrarily chosen origin point in an arbitrary direction, clockwise or anticlockwise, in which the coordinate monotonically increases from $0$ to $4$.
In this coordinate system, the points $A$, $B$ and $\Lambda$ each have a coordinate value $\alpha$, $\beta$ and $\lambda$, respectively, that is in the interval $\left[0,4\right)$.

We define a separation function between any two of these three points as follows.
Let $K,M\in\left\{A,B,\Lambda\right\}$ be two points on a 4-loop with coordinates $\kappa$ and $\mu$. Let $\abs{\mu-\kappa}$ be the absolute value of the difference between $\mu$ and $\kappa$. Define the separation $h\left(\kappa,\mu\right)$ between $K$ and $M$ as the length of the shortest loop section between $K$ and $M$:
\begin{equation}\label{Hdef}
	\begin{aligned}[b]
		h\left(\kappa,\mu\right) =\min\left(\abs{\mu-\kappa},4-\abs{\mu-\kappa}\right) 
	\end{aligned}
\end{equation}
For example, in Fig. \ref{fig:contourgreen}
\begin{equation}\label{Hloop1}
	\begin{aligned}[b]
		&h\left(\alpha,\beta\right)= 1.476 \\
		&h\left(\alpha,\lambda\right)= 0.742
	\end{aligned}
\end{equation}
In the top two Figs. \ref{fig:contourgreenAliceBoxed} and \ref{fig:contourgreenBobBoxed}
\begin{equation}\label{Hloop2}
	\begin{aligned}[b]
		&h\left(\beta,\lambda\right) = 1.782
	\end{aligned}
\end{equation}
while in the bottom two Figs. \ref{fig:contourgreenAlicef2-boxed} and \ref{fig:contourgreenBobBoxed2}
\begin{equation}\label{Hloop3}
	\begin{aligned}[b]
		&h\left(\beta,\lambda\right) = 0.733
	\end{aligned}
\end{equation}

In Sec. \ref{sec:graph} we have seen how for any node $a$ in the bipartite graph $G$, the weights $\{w\bigl(c_{ab_j}\bigr)|j \in \mathcal{L}\}$ are uniformly distributed in the interval $\left[0,2\right]$. Likewise, on a 4-loop, the lengths of the loop sections between a fixed point $A$ and a random point $B$ are uniformly distributed in the interval $\left[0,4\right)$. We can therefore, given a node $a_i$ in the graph, map all nodes $b_j$ in $G$ to points $B_j$ on half of the 4-loop such that the points $B_j$ are uniformly distributed on that half of the 4-loop. Thus, for any pair of key choices $\left(a_i,b_j\right)$, one can position a pair of points $\left(A,B\right)$ on a 4-loop such that the weight of the edge connecting $a_i$ and $b_j$ is equal to the separation between the points $A$ and $B$ on the 4-loop:
\begin{equation}
	\begin{aligned}\label{obsloop}
		w\bigl(c_{a_ib_j}\bigr)=h\left(\alpha,\beta\right)
	\end{aligned}
\end{equation} 
Also the uniform distributions of the green fluid levels $\lambda_A$ and $\lambda_B$ correspond to uniform distributions of the separations $h\left(\alpha,\lambda\right)$ and $h\left(\beta,\lambda\right)$, respectively.
\begin{equation}
	\begin{aligned}\label{GisH}
		\lambda_A= &h\left(\alpha,\lambda\right)\\
		\lambda_B= &h\left(\beta,\lambda\right)
	\end{aligned}
\end{equation}
Since the coordinate system around the 4-loop is merely a mathematical convenience, we can now choose the origin to be at point $A$, so $\alpha=0$, and choose the direction in which the coordinate monotonically increases such that the point $B$ has the coordinate $\beta=h\left(\alpha,\beta\right)=w\bigl(c_{a_ib_j}\bigr)$.

At this point it is time for an intermezzo that sheds more light on the important detail that makes the 4-loop in Fig. \ref{fig:contourgreen} different from the rectangle in Fig. \ref{fig:redblue}. While $\lambda_A$ and $\lambda_B$ can be changed freely and independently of each other by moving the points $A$ and $B$ with respect to $\Lambda$ along the 4-loop in Fig. \ref{fig:contourgreen}, the same $\lambda_A$ and $\lambda_B$ can clearly not move freely in the 2-dimensional plane in Fig. \ref{fig:redblue}.
With $\lambda$ uniformly distributed on $\left[0,4\right)$, the joint density of the pair of green fluid levels for a given pair of key choices is
\begin{equation}\label{rhoAB}
	\rho\left(\lambda_A,\lambda_B \mid a_i,b_j\right)
	=\frac{1}{4}\int_0^4 \delta\bigl(\lambda_A-h\left(0,\lambda\right)\bigr)\,
	\delta\Bigl(\lambda_B-h\bigl(w\bigl(c_{a_ib_j}\bigr),\lambda\bigr)\Bigr)\dd{\lambda}.
\end{equation}
This density is concentrated on the perimeter of the rectangle in Fig.~\ref{fig:redblue}, and it depends on both key choices through $w\bigl(c_{a_ib_j}\bigr)$. Its marginals, however, are uniform on $\left[0,2\right]$ for every key pair, so $\rho\left(\lambda_A\mid a_i,b_j\right)$ does not depend on $b_j$ and $\rho\left(\lambda_B\mid a_i,b_j\right)$ does not depend on $a_i$.
The statistical dependence of the pair of hidden variables $(\lambda_A,\lambda_B)$ on the settings $a_i,b_j$ violates statistical independence, which Sec.~\ref{Ch2} also calls no-conspiracy. It is not, however, a conspiracy in the sense of superdeterminism or retrocausality. Nothing in the model makes the choice of a key depend on $\lambda_A$ or $\lambda_B$: Alice and Bob are free to choose any key each time. Nor is any variable in the cross section of the past light cones of Alice and Bob's key choices tuned to those two key choices: $\lambda_A$ and $\lambda_B$ take their values when the yellow lights turn on. Each hidden variable on its own is independent of the distant setting: $\rho(\lambda_A\mid a_i,b_j)$ is uniform on $[0,2]$ for every $b_j$, and likewise for $\lambda_B$, so the model cannot be used to signal. The only thing that depends on both settings is the joint distribution of $\lambda_A$ and $\lambda_B$, see Eq.~\eqref{rhoAB}, and it depends on them only through the weight $w(c_{a_ib_j})$. Since $X=f(\lambda_A)$ and $Y=g(\lambda_B)$, that dependence is no more than the dependence of the observed correlation $\mathbb E_{a_ib_j}(XY)$ on the setting pair, which any account of the data must reproduce. There is an important distinction between my model and the trivial one in which the hidden variable simply records the settings and the outcomes. The trivial model can reproduce any correlations whatsoever, whereas in my model the hidden variables do not contain the settings and the correlations are constrained to the form $1-h(\alpha,\beta)$, where $h$ is a separation between random points on an $n$-sphere. In three dimensions ($n=2$), my model reaches Tsirelson's bound but cannot exceed it, while in two dimensions ($n=1$) it does not violate the CHSH inequality at all. It reads the coordination between $\lambda_A$ and $\lambda_B$ as a geometric relation between the two settings. Because nothing in their common past screens off that coordination, the model violates Bell's local causality (Sec.~\ref{sec:conclusion}), but, as in QM, does so in a way that cannot be used to signal: from her or his obtained data alone, neither Alice nor Bob can say anything about the other participant's key choices. The violation of local causality only shows up when the data of Alice and Bob are laid side by side, as a violation of the CHSH inequality. End of intermezzo.

An illustration will clarify what we have accomplished so far. In Fig. \ref{fig:contourgreen} we have $\alpha=0$ and $\beta=h\left(\alpha,\beta\right)=w\bigl(c_{a_ib_j}\bigr)=1.476$. The coordinate runs anticlockwise. In the upper two Figs. \ref{fig:contourgreenAliceBoxed} and \ref{fig:contourgreenBobBoxed}, $\lambda$ has a coordinate higher than $\beta$, namely $\lambda=\left(-\lambda_A\right) \bmod 4=\left(-0.742\right) \bmod 4=3.258$. In the lower two Figs. \ref{fig:contourgreenAlicef2-boxed} and \ref{fig:contourgreenBobBoxed2}, $\lambda$ is less than $\beta$, namely $\lambda=\lambda_A=0.742$. In all four figures, we have $f\left(\lambda_A\right) = 1 \text{ ($not{-}green$)}$
since $\lambda_A=h\left(0,3.258\right)=h\left(0,0.742\right)=0.742 < 1$. The value of $g\left(\lambda_B\right) = -1 \text{ ($\textit{green}$)}$ in the upper two figures, because $\lambda_B=h\left(1.476,3.258\right) = 1.782 > 1$. In the lower two figures, $g\left(\lambda_B\right) = 1 \text{ ($\textit{not-green}$)}$, since $\lambda_B=h\left(1.476,0.742\right) = 0.733 < 1$.

Let us consider the expectation value of $\mathbb E(f\bigl(\lambda_A\bigr)g\bigl(\lambda_B\bigr))$, with $\lambda$ uniform on $\left[0,4\right)$ as in Eq. \eqref{rhoAB}. This expectation value can take any value in the interval $\left[-1,1\right]$. We can compute $\mathbb E(f\bigl(\lambda_A\bigr)g\bigl(\lambda_B\bigr))$ by taking the average of infinitely many terms while varying $\lambda$ over a uniform discrete distribution in the interval $\left[0,4\right)$, as follows: 
\begin{equation}
	\begin{aligned}[b]\label{HV1}
		\mathbb E(f\bigl(\lambda_A\bigr)g\bigl(\lambda_B\bigr)) = & \lim_{N \to +\infty} \frac{1}{4N} \sum_{k=0}^{4N-1} f\left(h\left(\alpha,\lambda\left(\frac{k}{N}\right)\right)\right)g\left(h\left(\beta,\lambda\left(\frac{k}{N}\right)\right)\right)\\ 
		=&\frac{1}{4}\int_0^4 f\left(h\bigl(\alpha,\lambda\bigr)\right)g\left(h\bigl(\beta,\lambda\bigr)\right)\dd \lambda
	\end{aligned}
\end{equation}
In Eq. \eqref{HV1}, $\lambda\bigl(\frac{k}{N}\bigr)$ can be a random variable that has a uniform distribution in $\left[0,4\right)$ or e.g. $\frac{k}{N}$ itself.
We can express $\mathbb E(f\bigl(\lambda_A\bigr)g\bigl(\lambda_B\bigr))$ in terms of $h\left(\alpha,\beta\right)$ as follows. $f\left(h\bigl(\alpha,\lambda\bigr)\right) = 1$ on half of the 4-loop, namely, where $h\left(\alpha,\lambda\right) \le 1$. On the other half of the 4-loop, where $h\left(\alpha,\lambda\right) > 1$, $f\left(h\bigl(\alpha,\lambda\bigr)\right) = -1$. Similarly, $g\left(h\bigl(\beta,\lambda\bigr)\right) = 1$ on half of the 4-loop. On the other half of the 4-loop, $g\left(h\bigl(\beta,\lambda\bigr)\right) = -1$. The borders between the half loops are not the same for $A$ and $B$, but offset from each other by the separation $h\left(\alpha,\beta\right)$ between $A$ and $B$. The two sections where $f\left(h\bigl(\alpha,\lambda\bigr)\right) = -g\left(h\bigl(\beta,\lambda\bigr)\right)$ (the sections marked blue in Fig. \ref{fig:contourgreen}) together have a length $2 \times h\left(\alpha,\beta\right)$. The two sections where $f\left(h\bigl(\alpha,\lambda\bigr)\right) = g\left(h\bigl(\beta,\lambda\bigr)\right)$ (the sections marked red in Fig. \ref{fig:contourgreen}) together have a length $4 - 2 \times h\left(\alpha,\beta\right)$. The contribution to the integral from the two sections of the 4-loop where $f\left(h\bigl(\alpha,\lambda\bigr)\right) = -g\left(h\bigl(\beta,\lambda\bigr)\right)$ is $-1 \times\left[2 \times h\left(\alpha,\beta\right)\right]$. The contribution from the two loop sections for which $f\left(h\bigl(\alpha,\lambda\bigr)\right) = g\left(h\bigl(\beta,\lambda\bigr)\right)$ is $1 \times\left[4 - 2 \times h\left(\alpha,\beta\right)\right]$. Adding all contributions together and dividing by the perimeter of the 4-loop, we obtain 
\begin{equation}
	\begin{aligned}[b]\label{Prt}
		\mathbb E(f\bigl(\lambda_A\bigr)g\bigl(\lambda_B\bigr)) =\frac{-1 \times\left[2 \times h\left(\alpha,\beta\right)\right] + 1 \times\left[4 - 2 \times h\left(\alpha,\beta\right)\right]}{4} = 1-h\left(\alpha,\beta\right)
	\end{aligned}
\end{equation}
According to Eq. \eqref{obsloop}, it then follows that Eq. \eqref{Prt2} is true. 
\begin{equation}
	\begin{aligned}[b]\label{Prt2}
		\mathbb E(f\bigl(\lambda_A\bigr)g\bigl(\lambda_B\bigr)) = 1-w\bigl(c_{ab}\bigr)
	\end{aligned}
\end{equation}
From the results of the first experiment, where Alice and Bob made observations with binary (\textit{green}, \textit{not-green}) outcomes, we already know the weight of each edge $c_{a_ib_j}$, because it is computed from the correlation between the colour observations.
\begin{equation}
	\begin{aligned}\label{correlation}
		w\bigl(c_{a_ib_j}\bigr) =1 - \mathbb E_{a_i b_j}(X Y) 		
	\end{aligned}
\end{equation}
Combining Eqs. \eqref{correlation} and \eqref{Prt2} yields Eq. \eqref{Prt4}.
\begin{equation}
	\begin{aligned}[b]\label{Prt4}
		\mathbb E_{a_i b_j}(X Y) = \mathbb E(f\bigl(\lambda_A\bigr)g\bigl(\lambda_B\bigr))
	\end{aligned}
\end{equation}

If Alice's and Bob's green fluid levels, $h\left(\alpha,\lambda\right)$ and $h\left(\beta,\lambda\right)$, have uniform distributions in $\left[0,2\right]$ and are also pairwise independent (that is, if Alice's and Bob's key choices are not kept fixed), then $h\left(\alpha,\beta\right)$ cannot be fixed but must also have a uniform distribution in $\left[0,2\right]$. With Eq. \eqref{Prt}, it follows that $\mathbb E(f\bigl(\lambda_A\bigr)g\bigl(\lambda_B\bigr))$ has values in the interval $\left[-1,1\right]$ and that it is uniform in that interval.

In the case of three spatial dimensions we have, according to Eq. \eqref{H2}
\begin{equation}
	\begin{aligned}\label{pqcosangleB}
		h\left(\alpha,\beta\right) =\leftidx{_2}{H}
		{}\left(\eta_{\hat{a}\hat{b}}\right)=1 - \cos(\eta_{\hat{a}\hat{b}})
	\end{aligned}
\end{equation}
Finally, Eqs. \eqref{pqcosangleB} and \eqref{Prt} lead to Eq. \eqref{pqangle}.
\begin{equation}
	\begin{aligned}\label{pqangle}
		\mathbb E(f\bigl(\lambda_A\bigr)g\bigl(\lambda_B\bigr)) = \cos(\eta_{\hat{a}\hat{b}})
	\end{aligned}
\end{equation}
Eq.~\eqref{correlation} might suggest that the model simply fits the weights to the observed correlations. That is not the case. The weights are not free parameters: according to Eq.~\eqref{Eisw}, each weight is a function of the closeness angle between two random points on a sphere, and it is this geometry, not the data, that limits the correlations the model can produce. In three spatial dimensions the geometry gives Eq.~\eqref{pqangle}, so ${\left|S\right|}$ cannot exceed Tsirelson's bound, and in two spatial dimensions the CHSH inequality is not violated at all. Eq.~\eqref{correlation} only identifies which weight belongs to which pair of keys.

\subsubsection{Wrapping up}\label{sec:conclusion}
The original goal of hidden variables theories was to account for the values of measurable physical quantities irrespective of whether such quantities are, in fact, measured. Such a theory would have greater predictive and explanatory power than QM, which does not go further than predicting the probability of obtaining a specific outcome given the measurement setup.
I reject counterfactual definiteness; I have introduced a mathematical construction, ``the 4-loop'', that accounts for the actual outcomes in each single trial, not for values of incommensurable quantities that could have been measured instead.
Here are the main points:
\begin{itemize}
	\item The joint distribution of $\lambda_A$ and $\lambda_B$ depends on both key choices, see Eq.~\eqref{rhoAB}. The model therefore does not satisfy the statistical independence condition of Sec.~\ref{Ch2}. 
\item There is no setting-independent variable from which $\lambda_A$ follows using Alice's key alone and $\lambda_B$ using Bob's key alone; if there were, the model would satisfy the CHSH inequality.
	\item Not all of Bell's assumptions underlying his proof can be true.
	Given $(\lambda_A,\lambda_B)$ the outcomes factorise, and each marginal is independent of the distant key, so the model cannot signal. This factorisation is not Bell's local causality, because the joint distribution of $(\lambda_A,\lambda_B)$ depends on both keys: statistical independence fails. Local causality asks whether the correlation between $\lambda_A$ and $\lambda_B$ is screened off by a setting-independent state in their common past. It cannot be, for then the model would satisfy the CHSH inequality (previous bullet). The only ways left to keep local causality are superdeterminism and retrocausality, which I reject. The failure of statistical independence and the failure of local causality are therefore one fact, not two: the model, like QM, violates Bell's local causality while respecting no-signalling.
	\item The goal of the 4-loop is to explain, by analogy with Carol's experiment, how the outcomes of a pair of actual spin component measurements on an entangled state are the work of a pair of hidden variables $\lambda_A$ and $\lambda_B$.
	\item The hidden variables do not accommodate the simultaneous determination of properties that are incommensurable. They only accommodate actually measured properties.
	\item The equation that expresses how the expectation value of the product of Alice and Bob's outcomes depends on the hidden variables has, for each fixed key pair, a formal resemblance to the formula that, according to Bell, applies to local hidden variable theories, except that $\rho$ depends on the key pair.
	\item The presented model can violate the CHSH inequality ${\left|S\right|} \le 2$.
	In three spatial dimensions, ${\left|S\right|}_{\max}$ is Tsirelson's bound, $2\sqrt{2}$. In higher dimensions, ${\left|S\right|}_{\max}$ approaches the superquantum correlation \citep{Popescu1994QuantumNA} limit of $4$. In two spatial dimensions the geometry gives $\left|S\right|_{\max}\le 2$, so the CHSH inequality is not violated.
	\item The 4-loop is indifferent to the actual number of spatial dimensions. A supplementary physical theory is needed to provide the actual number of spatial dimensions.
	\item There is a caveat. The points above take the hidden variables at face value. The 4-loop proposition does not state whether the hidden variables are real or just the product of post-hoc rationalisation, analogous to the application of realistic-looking colours to a Daguerreotype. In its current state, the model does not propose a physical mechanism by which the apparatuses could work as described. Rather it invites us to think about how we would build a physical theory to explain the results of Carol's experiment, if it were ever realised.
\end{itemize}

\section{Discussion}\label{discussion}

\subsection{Richard's view of Inge's and Bart's positions}\label{sec6.1}

\subsubsection{On Inge}
I find Inge's position rather puzzling.
I have the impression that he hasn't fully understood the derivation in Section~\ref{Ch2} which started with the assumption which Bell calls ``local causality'', and derived from that a joint probability distribution of counterfactual variables $X_1$, $X_2$, $Y_1$ and $Y_2$, statistically independent of the settings $A$, $B$. What this means precisely is that we start with a joint distribution of $A$, $B$, $X$ and $Y$ and then augment the probability space on which they are defined so as to include additional variables $X_1$, $X_2$, $Y_1$ and $Y_2$ such that $X = X_A$ and $Y = Y_B$ which reflect classical spatio-temporal causality. 

By the classical causality assumptions, $(A, B)$ is statistically independent of $(X_1, X_2, Y_1, Y_2)$. If one thinks of the settings as being fixed values $a$, $b$, or as one could say, if they are ordinary mathematical variables, then the construction ensures that the probability distribution of the four-tuple $(X_1, X_2, Y_1, Y_2)$ is independent of the settings $(a, b)$. 

Note that there is no assumption that measurement outcomes are the values of predetermined and preexisting variables, revealed by measurement. It is not an ontological assumption. It is an epistemological assumption: the joint probability distribution of what we can observe can be thought of as built up in a particular way. 
Naturally, that \emph{enables} certain ontological interpretations. That is a matter of philosophy and the foundations of science. From the statistician's point of view, the probability distributions determined by quantum mechanics are probability models for a repeatable experiment; the probabilities correspond to long-run relative frequencies. A Bayesian statistician may want to interpret the probabilities subjectively. He or she expects the corresponding relative frequencies to turn up in repeated experiments.

Let me point out that the first part of this exercise is always mathematically possible, and indeed, in a completely trivial way! For instance, define $X_1 = X$ if $A =1$, and, for instance, $X_1 = +1$ if $A = 2$; and so on. Three more arbitrary choices are needed in order to define the remaining three variables. However, this particular augmentation, which does not require the underlying sample space to be augmented, will usually not satisfy the independence condition.

In fact, it is a theorem due to \citet{PhysRevLett.48.291} that the full augmentation is possible if and only if the distribution of $(X, Y, A, B)$ satisfies all no-signalling equalities and all one-sided CHSH inequalities; or if one prefers: when the four conditional distributions of $(X, Y)$ given $(a, b)$ satisfy those same constraints.

Inge writes ``I do believe in the thesis that $(A, B)$ is independent of what in Section~\ref{Ch2} is called $(\lambda_L, \lambda_X, \lambda_Y)$''.
But these three random variables arise as a
mathematical trick, in the context of the Markovian assumptions portrayed graphically in the DAG of Figure \ref{fig2}, which encodes a collection of conditional independence
assumptions involving random variables associated with each node. Nodes $A$, $B$, $X$, and $Y$ are
associated with random variables with the same name; nodes $E$ and $L$ are associated with hidden variables $\lambda_E$ and $\lambda_L$. Node $E$ allows $(A, B)$ to have an arbitrary joint distribution,
so in particular it could be a pair of fixed values $a, b$. Node $L$ takes care of statistical dependence
between the two measurement outcomes $X$ and $Y$ conditional on the two settings $a$, $b$.
In particular, it expresses that conditional on the random variables associated with 
nodes $A$, $B$, $E$ and $L$, the random variables $X$ and $Y$ are statistically independent of one another,
and the conditional distribution of $X$ depends only on $A$ and $\lambda_L$.
Therefore we can represent the conditional probability distribution
of $X$ given $A$ and $\lambda_L$ by thinking of $X$ as a deterministic function of $a$, $\lambda_L$ and 
of an auxiliary random variable $\lambda_X$, which might as well be taken to be uniformly distributed on the interval $[0, 1]$. Each auxiliary variable introduced this way is statistically independent of the others.

He goes on to write ``I will argue below that the variables $X_1$, $X_2$ , $Y_1$ and $Y_2$ may not have a joint distribution, so then $\lambda_X$ and $\lambda_Y$ seem to be defined on different probability spaces''. He is putting the cart before the horse. If we assume local causality, then all these mathematical objects can be defined as random variables on one probability space on which $A$, $B$, $X$, and $Y$ are also defined, and where statistical independence holds. From this, the CHSH inequalities follow. But Inge does not start with local causality. He starts with some assumptions inspired by quantum theory. He then derives more of quantum theory. He then finds that the CHSH inequality can be violated. So he may safely forget the first half of Section~\ref{Ch2}.

As I see it, Inge's programme is to select some key facts from quantum mechanics, and to derive the rest from that. In his theory he will only talk about a joint probability distribution of variables for which QM actually allows joint measurement. When he talks of persons having variables in their mind, he is talking about predictive probabilities in a future experiment in which exactly and only those variables could be observed. If we classify variables according to his accessibility criterion, we will only be interested in joint distributions of variables which can be observed simultaneously. Thus the non-existence of joint probability distributions for experiments which the theory does not talk about, is no surprise. There is absolutely no need for such to exist. 

The part of quantum theory which Inge recovers is the core part with a separable complex Hilbert space. It can be finite dimensional, or infinite dimensional. He treats those two cases separately (different groups are involved).

Now, in Section \ref{Inge} he pays little attention to the Born rule, which is the part of QM which actually predicts experimental results -- long-run frequencies in many repetitions. He does derive certain probability distributions for particular states and measurements from symmetry considerations. That is nice, and part of a long tradition. But not all situations allow those symmetries; and anyway, experimental data might show they are not true! So I would say that this is a very meagre part of QM as it exists today.

Regarding the Born rule, Inge refers to recent work in which he uses the likelihood principle, or sufficiency, to derive the Born rule. On inspection it turns out that he is using a fact from quantum theory, which can be verified by experiment. Let me talk about it in the 2-dimensional case. We measure the spin of a particle in some pure state (corresponding to a direction), and in some direction. We get a binary outcome -- spin-up or spin-down in the measured direction. The von Neumann-L\"uders projection postulate says that the particle is now in the corresponding pure state. A subsequent measurement of spin in a different direction gives again $\pm 1$ and the probabilities of those outcomes depend only on the angle between the two directions. Repeat again, measuring in another direction. The same happens again. This is the important fact: that whole process is Markovian: the probabilities of future measurement outcomes only depend on the present situation, and on chosen measurement settings, but not on the past history.

A priori, it need not have been true. It can be observed to be true in certain situations. 

Thus Inge has focussed on a small number of important features of some rather special models in quantum theory, and has derived the remaining parts of those specific models. The Hilbert space, some special states, some special measurements.

I find the work important and very interesting, but I do not see that it helps to understand quantum theory. The properties which he takes as central are not intuitive to me.
They can however become intuitive after one has got accustomed to quantum theory!

\subsubsection{On Bart}

I find Bart's contribution very puzzling, too. Part of the problem is that he somewhat abruptly transitions from
a visual picture of an experiment in which Alice and Bob each select settings from very large collections of settings, to a particular case in which Alice and Bob's settings are both fixed. Like Inge, he does not treat the four sub-experiments of a CHSH experiment as four parts of one joint model. He does not have an explicit mathematical model till the end of his section, in which he introduces the ``4-loop'', as a model for how two measurement outcomes arise from two given settings. But he does not provide a mechanism explaining how that model could arise.

Bart regards the rejection of counterfactual definiteness as the basis for a hidden-variable model that recovers the violation of the CHSH inequality. I will argue that his final model is a \textit{non-local} hidden variable model, which makes violation of the CHSH inequality easily possible. It seems to me that the desired violation is inserted ``by hand''. 

He starts with a long description of a new experiment, called Carol's experiment, in which Alice and Bob each make choices from very many different settings. The model description talks about variables $\lambda_A$ and $\lambda_B$ but postpones a specification of a joint probability distribution
of these two variables, or a specification of a physical mechanism which explains how they are realised. 
Their existence is simply assumed. At the end of a long exposition he transitions to a very specific probability model, which he calls the 4-loop, which describes how the joint probability distribution of the two outcomes can be expressed by seeing them as a function of one uniformly distributed random variable $\Lambda$. 

This model is completely explicit. It uses the experiment of Carol in order to \emph{geometrically} specify a certain function $h(\alpha, \beta)$ and from it, $\lambda_A$ and $\lambda_B$. The model requires specification of a dimension $n$. In the case of interest $n=2$; Alice and Bob's settings are arbitrary directions in ordinary 3-dimensional space, hence represented by points $a$, $b$ on the sphere $S^2$. But what are $\alpha$ and $\beta$? Equation (\ref{GisH}) defines $\lambda_A$ and $\lambda_B$ in terms of $\alpha$, $\beta$ and $\lambda$; not in terms of $a$ and $b$.

For $n = 1, 2, \dots$, Bart introduces successively functions $\eta$, $w$, $I$, ${}_n H$, then gives further information about $w$, and finally $h$, the separation between two points on the 4-loop (Eq.~\eqref{Hdef}).

From now on, I will focus on the case $n =2$. For an angle $\nu$, 
define ${}_2 H (\nu)$ to be twice the proportion of the area of the 2-sphere within an angle $\nu$ of the apex. It runs from $0$ to $2$ as $\nu$ runs from $0$ to $\pi$. By trigonometry, ${}_2 H (\nu) = 1 -\cos(\nu)$.

Pick two settings $a$, $b$, points in $S^2$. We compute the number $h = w(c_{a, b}) = {}_2 H (\eta (a,b))$. Now we place two special points $A$, $B$ on the 4-loop at distance $h$ from one another. Next, a point called $\Lambda$ is chosen completely at random on the irregular ``4-loop'' in Figure \ref{fig:contourgreen}. 

The three locations on the 4-loop called $A, B, \Lambda$ are parameterised by coordinates $\alpha, \beta, \lambda \in [0, 4)$. The settings on the other hand are directions $a$, $b$ in 3-d space. This is a hidden variable theory for one component of a Bell experiment, from which no conclusions can be drawn at all. Given two pairs of settings $a, a'$ and $b, b'$, we can consider the four sub-experiments of a CHSH experiment. Each can be modelled separately with a hidden variables model. Can they be modelled jointly in one hidden variables model?

\begin{figure}
	\noindent\centerline{\includegraphics[width=0.4\textwidth]{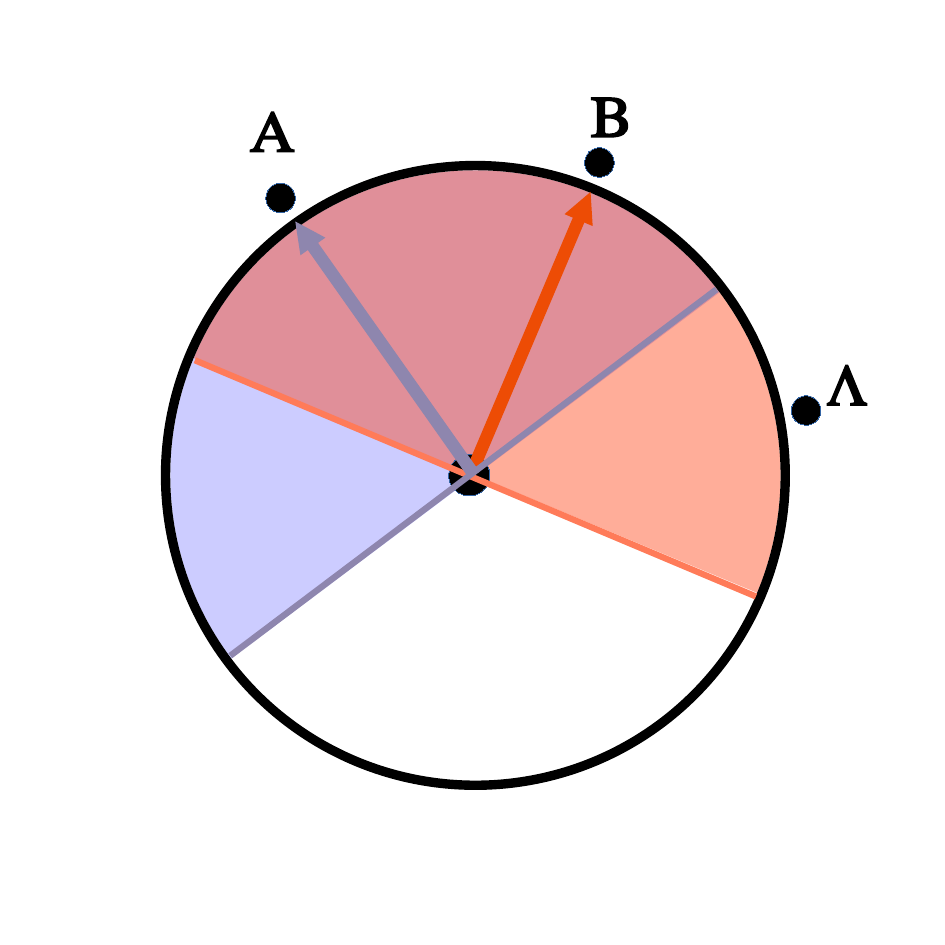}}
	\caption{Bart's model, the 4-loop has become a circle of circumference 4. Imagine two concentric perspex disks. One half of each disk is fully transparent, the other half is semi-transparent. On each disk an arrow is printed pointing from the centre to the ``summit'' of the semicircular boundary of the coloured half. These are the
		points labelled $A$ and $B$. In this colour illustration, the two colours are a pale violet and a pale brick-red. In Bart's construction, first the distance between $A$ and $B$ is determined, which can be done either by
		a Monte Carlo experiment involving random points selected on an $n$-sphere, or calculated analytically. $A$ and $B$ are placed on the 4-loop at this distance between one another, and independently of that, a point $\Lambda$ is chosen uniformly at random on the 4-loop.}
	\label{fig9}
\end{figure}

We may as well visualise the 4-loop as a circle with perimeter of length $4$, thus radius $2/\pi$, see Figure \ref{fig9}. Given a setting pair $a, b$, place two points on the circumference called $A$ and $B$ at the minimal distance $h(a, b)$ measured clockwise, or anticlockwise, between $A$ and $B$. Thus $h \in [0, 2]$. Parameterise the circle by points in the interval $[0, 4]$ with endpoints identified, and arithmetic done modulo 4. Without loss of generality $A$ can be located at the point with parameter $\alpha = 0$, so that $B$ is at the point with parameter $\beta = \pm h$. Now pick a random number $\lambda$ uniformly distributed on $[0, 4)$ and take it to be the parameter of the random point which Bart calls $\Lambda$.

Each of $\alpha$ and $\beta$ can be thought of as the midpoint of a segment of length $2$, i.e., a semicircle. Intersecting those two semicircles or their complements, we generate four disjoint segments of the circle. When we pick the parameter $\lambda$ uniformly at random from $[0, 4)$ and check which subsegment the point $\Lambda$ with coordinate $\lambda$ lives in, we generate two binary outcomes, with joint probability mass function taking four values $1/4\pm c/4$ such that marginally, we have two fair Bernoulli trials; and the correlation is $c$. Depending on $h$, all correlations in $[-1, +1]$ are possible.

The distance between the point called $A$ to the point called $B$ is explicitly defined as a function of the two settings $a$ and $b$. We need to know the settings in advance, before we can run a simulation of the core part of his model. The \emph{modelling} is non-local.

Bart's own Figure \ref{fig:redblue} makes the same point in another way. The pairs of green fluid levels $(\lambda_A, \lambda_B)$ lie on the perimeter of a rectangle whose shape depends on the actual pair of key choices. Bart seems to take $\lambda_A$ to represent how Alice `sees' the shared variable $\lambda$ from the perspective of her setting, and $\lambda_B$ is how Bob sees it from the perspective of his. But how Alice's perspective is placed relative to Bob's is fixed by \emph{both} key choices. There are two ways to read this. If we take the pair $(\lambda_A, \lambda_B)$ to be the hidden variable, then its probability distribution depends on the settings $(a, b)$, and statistical independence is violated. If we take the point $\Lambda$ on the loop to be the hidden variable (Bart explicitly rejects this interpretation) then Alice's outcome depends only on $\lambda$, but Bob's outcome depends on $\lambda$ and on $\beta = w\bigl(c_{ab}\bigr)$, which depends on Alice's key choice: locality is violated. Either way, one of the assumptions of Section~\ref{Ch2} fails. 
Whatever the value of $n$, the resulting model satisfies all four no-signalling equalities. It satisfies the 8 one-sided CHSH inequalities in the case $n = 1$ but not in the case $n = 2$ (or higher). By Fine's converse of Bell's theorem, it can be represented in the Bell form $X = F(A, \lambda)$, $Y = G(B, \lambda)$ with $\lambda$ independent of $(A,B)$ when $n = 1$ but not when $n = 2$ (the case of most interest).

It evidently can be written in the form $x = F(\lambda; a, b)$ and $y = G(\lambda; a, b)$. We proved in Section~\ref{Ch2} that it cannot be written in the form $x = F(\lambda; a)$ and $y = G(\lambda; b)$
with $\lambda$ independent of $(a,b)$. That proof does not use coordinates or a given frame of reference, only the spacelike separation of the two wings of the experiment.

The problem is that if we try to use Bart's model to describe a CHSH experiment, we will need to consider directions $a, a', b, b'$. Thus we will have four loops of four with points $(A, B, \Lambda)$, $(A', B, \Lambda)$,
$(A', B', \Lambda)$, $(A, B', \Lambda)$. For a violation of CHSH we will not be able to merge those four subexperiments to one by having the same position of $A$ both in company of $B$ and $B'$, etc.

My conclusion: Bart has come up with a complicated-looking but actually very simple model. It is indeed important to distinguish between the names of the settings chosen by Alice and Bob and the actual settings implemented inside the measurement device. In real experiments, a long period of calibration takes place before one trial in a final Bell test. The local coordinate systems in the two labs need to be ``correlated'' with one another. Bart's model reveals hidden subtleties in Bell's theorem, which is popularly presented in a simple way in which important assumptions are being made implicitly. It is silently assumed that the space of possible values of the hidden variable does not depend on the settings $a$ and $b$, but in Bart's model this is not the case. The space comes with special points whose location depends on $a$ and $b$. Bart has cunningly encoded the distance between $\alpha$ and $\beta$ in the space of values of $\lambda$. That space isn't $[0, 4)$ but $[0, 4)$ with some special points marked with labels $a$, $b$.

The model therefore has important didactic value. 

\subsection{Inge's view of Richard's and Bart's positions}\label{sec6.2}

\subsubsection{On Richard} 

I appreciate Richard's contribution. He has deep insights into quantum mechanics, and this, combined with his statistical background, has given him a unique position in defending the Bell theorem, and also in describing the Bell theorem in agreement with all the writings of John Bell. He is also well-informed on the necessary physical literature and on the Bell experiments that have been done so far. All this insight is reflected in his contribution.

Richard and I agree on the mathematics of Section~\ref{Ch2} and on the rejection of superdeterminism and retrocausality. We also agree to reject explanations based on non-locality in the form of action at a distance and explanations given by Bell-deniers. Furthermore, we agree to reject mechanical explanations: There does not exist a local, classical mechanism that `explains' the observations given by quantum mechanics.

Where we part company is what to do next. Richard is content to call the randomness irreducible, to treat quantum probabilities and classical probabilities as the same kind of number differing only in origin, and to leave the ontological question to the philosophers. I find this too limited. A probability that is not attached to a variable, and a variable that is not attached to an observer who can in principle access it, leave me without a place to stand. I would rather say: Locality in the form of no action at a distance we keep; realism, in the sense of a context-independent joint value for non-commuting observables, we give up; and the formalism that replaces realism can in the Bell experiment situation be derived from some plausible postulates. Richard's position is consistent, but it stops one step before the step I consider the interesting one.

However, as mathematical statisticians we agree on much. First, I fully support Richard's statement (Sec.~\ref{Richard}) that quantum mechanics has no universal set $\Omega$ on which observables are simultaneously defined, and no dispersion-free states, cf. \citet{gill2022belavkin}. Next, we both support George Box's aphorism that all models are wrong but some are useful. 

One question on which we seem to differ, is on the solution of the Schr\"{o}dinger's cat problem. To me, this problem can be solved by rejecting the universal validity of the superposition principle: All valid state vectors should be eigenvectors of some physically meaningful operator. Thus, from this point of view, I do not need the deep theory of Belavkin. But of course, I accept that Richard refers to and uses this theory.

I agree that, if there is a Heisenberg cut, it should be between the past and the future. But I prefer a solution where the role of the Heisenberg cut is limited. To me, quantum mechanics is a ubiquitous model, also valid for macroscopic systems. Classical mechanics is only a good approximation. So, in particular, QM is also a valid model for the measurement apparatus and for our own minds when we receive measurement results.

Belavkin's mathematical approach rests on the concept of operators as representing observables; these may either be beables (real observables connected to the past or present) or predictables (belonging to the future) see \citep{gill2022belavkin}. In my own approach, I give conditions under which accessible theoretical variables may be represented by such operators. This gives a theory where we can `explain' the violations of the CHSH inequality by assuming that all observers are limited in a certain sense. Such direct explanations may seem to be missing in the approach by Belavkin, as advocated by Richard.

However, I agree with Belavkin's construction of a Markov process describing past and future position of a particle. At least, I think that this construction is valid in the non-relativistic version of quantum theory. If confronted with general relativity theory, I think that the assumed universality of the `now' notion may be discussed. According to relativity theory, the `now', the notion of simultaneity, the cut between past and future, depends on the observer's frame of reference. To me, this is another argument for introducing an observer in quantum theory.

\subsubsection{On Bart}
Bart's Carol experiment is an extension of the Bell experiment: Alice has now $k$ settings to choose between, while Bob has $l$ settings (keys) to choose between. In the illustration and simulation, $k=l=30$. The key choices are labelled $A$ and $B$, while the observed responses (green or not green) are labelled $X$ and $Y$. The connection to the Bell experiment is then given by $\mathbb{E}(X_a Y_b) = \mathbb{E}(XY|A=a, B=b)$ for various choices of $a$ and $b$. As in the Bell experiment, the basic variables of the Carol experiment are $(A,B,X,Y)$.

A mathematical model for the Carol experiment is described on two levels. The first uses random weights $w\bigl(c_{ab}\bigr)$, constructed geometrically from the surface of an (hyper)sphere, where three spatial dimensions ($n=2$) correspond to the ordinary Bell case. Here the sphere is the familiar Bloch sphere. This could be compared to my own theory, where I need two Bloch spheres for modelling the two outputs.

In the second level of the model, the colours $X$ and $Y$ are functions of `hidden' fluid values $\lambda_A$ and $\lambda_B$. By simulation from the first model, it is shown that, for given $A=a$ and $B=b$, the points $(\lambda_a,\lambda_b)$ are distributed on rectangles embedded in the square $[0,2]\times[0,2]$. Bart calls a modified version of this rectangle the 4-loop.

The 4-loop makes it less visible that different settings by Alice and Bob produce different rectangles.

In these simulations, Bart finds values of the CHSH quantity $S$ that correspond to the values from quantum mechanics. In particular, the CHSH inequality turns out to be violated for certain choices of settings, and the maximum of $|S|$ is given by the Tsirelson bound.

Can this model `explain' why the CHSH inequality can be violated in practice? The model is interesting and ingenious, but in my opinion, the link to the Bell experiment is more direct when one concentrates on Figure \ref{fig:redblue} in Bart's contribution, not when the rectangles are reduced to 4-loops.

In the second level of the model, there are two `hidden' variables. This can be reduced to one by considering the circumference of the rectangle, but this is a kind of model reduction depending upon data. The rectangle is different for different choices of settings.

The whole discussion illustrates the position of realism without assuming counterfactual definiteness, a position that seems to be possible in relation to the Bell experiment, but I prefer to be more explicit and reject strict realism.

Bart has made extensive use of Claude in correcting errors in his contribution. This has increased my trust that details of his theory are correct. But current AI is not perfect. Claude may also have overlooked certain issues.

\subsection{Bart's view of Richard's and Inge's positions}\label{sec6.3}
\subsubsection{On Richard}\label{BartsViewOnRichard} Richard finds comfort in Belavkin's theory, which uses the formalism of unitary evolution of QM states to create a history that is fixed in the past and probabilistic in the future. The transition from future to past is irreducibly random. 

Central to Richard's rejection of local hidden variables is the widely held conclusion that Bell's local causality \citep{Bell_Aspect_2004_24} is excluded by the results of Bell-type experiments: no locally causal theory can reproduce those results. I share this view.

Richard is at ease with irreducible randomness in quantum mechanics. I think it is too early to accept that there are events in nature that just happen. It is impossible for me to contemplate this randomness as part of reality, independent of us, because, by definition, it does not work in any describable way. In addition, it is not clear how irreducible randomness can give comfort with the Bell proof. 
In QM, the entangled state coordinates the two random outcomes, but the state is not a mechanism: it does not say how two irreducibly random events come out correlated.
Doesn't Belavkin's theory imply a fixed trace backwards in time from the moment Alice does a measurement, and likewise for Bob? And aren't such traces hidden variables?

Richard objects that in my model the distance between $A$ and $B$ depends on both settings.
That is true: in my model, what correlates the two measurement results is the angle between the settings, together with the point on the 4-loop where $\lambda_A$ and $\lambda_B$ meet. Also according to QM, to sample outcomes from the Born rule one needs both settings.
The difference lies elsewhere.
Quantum mechanics keeps a setting-independent state and lets the outcomes be correlated given that state;
my model lets the outcomes factorise given $(\lambda_A,\lambda_B)$ and lets the distribution of that pair depend on both settings.
Neither is locally causal, and neither can signal.
Richard also notes that the four 4-loops of a CHSH experiment cannot be merged into one. That is correct, and it is the point of the model: merging them would amount to counterfactual definiteness, which I reject, and by Fine's theorem \citep{PhysRevLett.48.291} no model that violates the CHSH inequality can be merged in this way. Nor is the violation inserted by hand: the weights follow from the areas of spherical caps, and the same rule gives no violation in two spatial dimensions ($n=1$).

According to Richard's interpretation of quantum mechanics, irreducible randomness goes together with a fundamental time asymmetry. This implies that irreducible randomness exerts physical effects while, being irreducible, it is not itself under any physical influence.
My model, by contrast, says nothing about time: what I have presented in this paper is not a complete physical theory. It describes snapshots of zero duration (the lines in David's notepad), so it says nothing about how the correlations are
established in space-time.

I hope that my model can be extended with a physical component. It is certain that such a theory cannot add a hidden variable in the sense of Bell, independent of the settings. As far as I can see, the only way to give the model physical content is to frame it as a theory of how the structure of space-time emerges from Bell-type experiments, a minute minority of which are directed by us, and the majority of which happen at every instant everywhere without our noticing them. In such a theory the structure we see around us -- including the labels and the angles between settings -- is like an interference pattern caused by spacelike separated events. The theory would be causal in the sense that every effect lies in the future light cone of its causes, the causes	being the observations by Alice and Bob and the effect in the intersection of the future light cones of the observations being a refreshing of the angle between the chosen settings. How the angle arises in the first place, and how a refreshed angle reaches both apparatuses in later trials, are open problems.

Whereas Richard accepts indeterminism and respects statistical independence, and pays with the inexplicability of single outcomes, my model determines each single outcome from $\lambda_A$ and $\lambda_B$, and pays with a violation of statistical independence. Both accounts require an unexplained stochastic element (in my model the uniform $\lambda$), in my view a \textit{deus ex machina}.

\subsubsection{On Inge}\label{BartsViewOnInge} Of the three authors, Inge is the one who answers `yes' to the question whether QM can be considered complete. Richard says `no' because QM is probabilistic, but does not seek hidden variables, and I hypothesise that the answer is `no' because I believe a viable hidden variable theory is not yet ruled out. Since the CHSH inequality is violated by nature, Inge finds comfort in keeping statistical independence, but rejecting the full accessibility of all four variables $X_1$, $X_2$, $Y_1$ and $Y_2$ at the same time. Inge's theory starts from the restrictions on which variables are accessible, where `accessible' remains mathematically undefined but can be understood epistemologically as denoting what a person can know and communicate to other people. I look for a materialistic solution in the geometry of a hidden variable and perhaps even in the number of dimensions of space. In Inge's view, the actual values of complementary variables cannot be in an observer's mind simultaneously. In my approach, the mutual exclusion of the measurement setups is reason enough to reject the assumption of counterfactual definiteness as untestable. I think both positions have merits.

Inge's approach assigns a prime role to observers' minds. In this approach, the arrow of time comes for free; it is inevitable for a conscious mind to sense time ``flowing'', to remember the past and to estimate the chance of things happening in the future. This approach is perhaps the ultimate and most pure consequence of the widely accepted position that quantum mechanics is not a description of how the world works, but a description of what is knowable about the world.

I do not reject the role of observer-centred argumentation in science, but I think it should only be used as a last resort in physics. As I have tried to show in Sec.~\ref{Bart}, it may still be too early.

\section{Conclusions}\label{conclusions}
Having devoted a paper partly to our disagreements, it is in this conclusion reasonable to record what we agree about. The following points are shared by all three authors, except were stated.
\begin{enumerate}
\item The mathematical content of Section~\ref{Ch2} is correct. The CHSH inequality is the conclusion of a theorem about any model in which the three assumptions of locality, realism (counterfactual definiteness), and no-conspiracy hold simultaneously.
All steps of the derivation are agreed upon by all of us. 
\item Local causality implies counterfactual definiteness and statistical independence.
\item The %
Bell experiments\footnote{Some groups state that the freedom-of-choice loophole is not and cannot be closed.} have shown that the CHSH inequality may be violated by the amount predicted by quantum mechanics. This violation can not be explained by any of the loopholes catalogued in Section \ref{Ch3}. At least one of the assumptions of Bell's theorem must therefore fail in nature.

\item Richard and Inge reject superdeterminism and retrocausality as possible explanations of this. They agree strongly that a theory that can `explain' anything by coordinating the past with the future explains nothing in the scientific sense. Purchasing agreement with experiment at the price of the basic methodology of science is too high a price. Bart's model violates statistical independence and is no-signalling but not locally causal, like QM. In his model, the joint distribution of the two hidden variables depends on both settings, while each hidden variable on its own is independent of the distant setting. Bart does not regard this as superdeterminism or retrocausality, because the settings are chosen freely and the model does not make them depend on the hidden variables. He focuses instead on the geometrical relation between the two settings, which is fixed by the setting pair and shows up in the joint distribution of the hidden variables.

\item We also reject Bell-deniers in all their variants. A mathematical theorem is a theorem, and this applies particularly to the Bell theorem. The experiments are the experiments; any route to comfort theorems with experiments must pay the cost somewhere. Pretending that there is no cost is not an available route.

\item We reject the claim that Bell's theorem has established non-locality in the sense of influence between outcomes that is faster than the speed of light. The no-signalling equalities forbid any operational use of such an influence. A non-locality that is both real and experimentally inert is not a gain worth the expense.

\item The remaining space of serious responses to the violation of the CHSH inequality in practice is smaller than the literature sometimes suggests. It is essentially the space in which one declines either the statistical independence or the realism assumption in one of several forms: by accepting irreducible randomness (Richard), by reconstructing the quantum formalism from a theory of accessibility and then confronting an epistemic interpretation of this theory with the deduced limitation of all observers (Inge), or by replacing the algorithmic notion of local hidden variable with a geometric one, at the price of statistical independence (Bart).

\item On these detailed explanations we do not agree, and we have not tried to force an agreement. We think the reader is better served by three stated positions, each developed to the point where its costs and its promises are visible, than by an artificial compromise that none of us would sign alone.
\end{enumerate}

What we would like the reader to take away from this is as follows. The Bell-CHSH violation is a fact with consequences to any future description of the world, not a piece of exotic news from a remote corner of quantum theory. The question is no longer whether this fact is real, but how one's picture of the world accommodates it. That question, as this paper has tried to show, admits more than one honest answer.

\section*{Acknowledgements}

We are grateful to everyone with whom discussions have led us to the standpoints that we have stated in this paper.

\section*{Competing interest declaration}

There are no competing interests behind this paper.

\section*{AI declaration}
The ideas, arguments and conclusions in this paper are the authors' own. We used Anthropic's Claude Opus 5.5 to check the text for inconsistencies and gaps, and to suggest rewordings and clarifications; some passages were adopted from or based on its suggestions. The authors reviewed every suggestion and take full responsibility for the content. 

\raggedright
\frenchspacing
\bibliographystyle{plainnat}
\bibliography{HeJoGi}

\end{document}